\documentclass[twocolumn,prd,floatfix,preprintnumbers,a4paper,nofootinbib,superscriptaddress,groupedaddress]{revtex4-1}
\usepackage{natbib}
\usepackage[english]{babel}
\usepackage[autostyle, english = american]{csquotes}
\MakeOuterQuote{"}
\usepackage[utf8]{inputenc}
\usepackage{amsmath,amssymb,amsfonts}
\usepackage{bm}
\usepackage{graphicx}
\usepackage{siunitx}
\usepackage{float}
\usepackage{makecell}
\usepackage{mathtools}
\usepackage{multirow}
\usepackage{booktabs}
\usepackage[caption=false]{subfig}
\usepackage{listings}
\usepackage[usenames,dvipsnames,table,xcdraw]{xcolor}
\usepackage[colorlinks, urlcolor=blue,citecolor=blue,pdfborder={0 0 0}]{hyperref}
\usepackage[nameinlink,capitalise]{cleveref}
\usepackage{xspace}
\usepackage{array}

\newcommand{\nurates}{\texttt{bns\_nurates}\xspace}
\newcommand{\athenak}{\texttt{AthenaK}\xspace}
\newcommand{\nue}{\nu_e}
\newcommand{\anue}{\bar{\nu}_e}
\newcommand{\nux}{\nu_x}
\newcommand{\anux}{\bar{\nu}_x}
\newcommand{\Msun}{M_\odot}
\newcommand{\Ye}{Y_e}
\newcommand{\Nq}{N_{\rm q}}      % quadrature order

\begin{document}

% Suggestion: Accelerating Neutrino Opacities in Neutron Star Merger Simulations Using Artificial Neural Networks
\title{A Hybrid Neural Network Emulator for M1 Neutrino Reaction Rates in Binary Neutron Star Merger Simulations}

\author{Yi Qiu$^{1,2}$, David Radice$^{1,2,3}$, Raffaele Ferrari$^{4,5}$, and Albino Perego$^{4,5}$}

\affiliation{$^1$Institute for Gravitation and the Cosmos, The Pennsylvania State University, University Park PA 16802, USA}
\affiliation{$^2$Department of Physics, The Pennsylvania State University, University Park PA 16802, USA}
\affiliation{$^3$Department of Astronomy \& Astrophysics, The Pennsylvania State University, University Park PA 16802, USA}
\affiliation{$^4$Dipartimento di Fisica, Universit\`a di Trento, via Sommarive 14, 38123 Trento, Italy}
\affiliation{$^5$INFN-TIFPA, Trento Institute for Fundamental Physics and Applications, via Sommarive 14, 38123 Trento, Italy}

\date{\today}

\begin{abstract}
We present a neural network emulator for the \nurates library, which computes neutrino reaction rates used in binary neutron star merger simulations. Following a hybrid strategy, we emulate only the computationally dominant two-dimensional reactions---pair processes, bremsstrahlung, and inelastic neutrino-electron scattering---while evaluating the inexpensive one-dimensional processes still with \nurates. We find that the emulator reproduces the opacities to within a few percent across the relevant thermodynamic conditions and accelerates their evaluation by roughly an order of magnitude. When deployed in a global \athenak binary neutron star merger simulation, it recovers the remnant structure, neutrino luminosities, and ejecta composition of a reference run using the full \nurates library, at approximately half the overall cost.
\end{abstract}

\maketitle

\section{Introduction} \label{sec:intro}

Binary neutron star (BNS) mergers are among the most energetic and informative events accessible to modern astrophysical observations~\cite{Baiotti:2016qnr}. The landmark event GW170817~\cite{LIGOScientific:2017vwq} simultaneously revealed gravitational wave (GW) emission, a short gamma-ray burst (GRB~170817A), and a kilonova (AT2017gfo)~\cite{LIGOScientific:2017zic,LIGOScientific:2017ync,Perego:2017wtu,Drout:2017ijr}, confirming BNS mergers both as central engines for relativistic jets and as sites of rapid neutron capture (r-process) nucleosynthesis, and opening a new chapter in multi-messenger astrophysics~\cite{Lattimer:1977igd,Metzger:2010sy,Wanajo:2014wha,Wu:2016pnw,Kasen:2017sxr,Radice:2018pdn,Shibata:2019wef,Radice:2020ddv}. The merger evolution is governed by coupled physical processes, including general relativistic magnetohydrodynamics and neutrino transport~\cite{Foucart:2022bth}. In particular, neutrinos play a decisive role in setting the electron fraction $\Ye$ of the outflows through their interactions with the surrounding matter, and in redistributing energy and momentum within the remnant~\cite{Qian:1996xt,Hoffman:1996aj,Martin:2017dhc,Qiu:2025kgy,Qiu:2025ybw}.

Several neutrino transport schemes are feasible in BNS merger simulations, ranging from leakage~\cite{Ruffert:1998qg,Rosswog:1998hy,Sekiguchi:2011zd} and moment-based prescriptions~\cite{Thorne:1981nvt, Shibata:2011kx, Cardall:2013kwa, Foucart:2016rxm, Sekiguchi:2015dma, Sekiguchi:2016bjd, Radice:2021jtw, Foucart:2024npn, Cheong:2024buu} to Monte Carlo methods~\cite{Foucart:2017mbt,Foucart:2022bth,Foucart:2025nub} and guided moments method~\cite{Izquierdo:2023fub}. In particular, the gray two-moment (M1) scheme, which evolves the energy-integrated zeroth and first moments of the neutrino distribution function and closes the resulting system of equations with an analytic closure, is widely used due to its balance between accuracy and computational cost.

In moment schemes, the coupling between matter and radiation is set by gray neutrino reaction rates. These rates are supplied by either pre-computed tables~\cite{OConnor:2014sgn,Ng:2023syk} or by libraries that evaluate reaction kernels on the fly. One such state-of-the-art library is \nurates~\cite{Chiesa:2024lnu}, which takes the local thermodynamic and radiation quantities to construct the neutrino distributions, integrates over energy and angle, and returns the gray emissivities and opacities used in global simulations. On-the-fly rate calculations offer several advantages, including the absence of interpolation errors and the possibility of taking into account the actual neutrino spectrum from the neutrino transport~\cite{Chiesa:2024lnu}. However, physical accuracy comes at the expense of higher computational cost and the need for real-time feasibility in actual simulations often forces a compromise on physical fidelity. \nurates models several of the demanding two-dimensional (two neutrino phase space) kernels: neutrino pair processes, nucleon-nucleon bremsstrahlung and the often-neglected neutrino-electron/positron scattering (NEPS), all of which contribute substantially to the heavy-lepton neutrino opacities~\cite{Cheong:2024cnb,Rath:2026vfr,Foucart:2026kpd}. Within widely used approximations, evaluating these kernels is far more expensive than the simpler beta processes, making them challenging to include efficiently in production simulations~\cite{Sekiguchi:2015dma,Foucart:2016rxm,Foucart:2024npn}.

In this paper, we explore an approach in which neural networks (NNs) are employed to accelerate the evaluation of the neutrino opacities. This is the first application of neural networks to modeling neutrino reaction rates, although related emulators have been developed for nuclear equations of state~\cite{MorawskiBejger2020}, for turbulence modeling~\cite{Karpov2022,Cook:2026tum}, for subgrid modeling of small-scale feedback on large-scale dynamics~\cite{Bhojwani:2025rtn}, and for neutrino flavor oscillations~\cite{Abbar:2023kta,Richers:2024zit,Chakraborty:2026vhk}. The central challenge in applying NN emulators to neutrino opacities is the multi-scale nature of the problem: the opacity varies over many orders of magnitude across the thermodynamic phase space of BNS conditions, with sharp gradients near the extreme cases~\cite{Thummler:2026baz}. Constructing an appropriate training set and efficient network architecture is therefore both challenging and decisive for performance.

To meet this challenge, we adopt a strategy whereby NNs are used only to emulate the computationally dominant processes, while we compute the less-expensive processes with \nurates. We show that the resulting emulator reproduces the full-library opacities typically to within a few percent across the relevant thermodynamic conditions, accelerates the opacity evaluation by roughly an order of magnitude, and, when deployed in a live \athenak~\cite{Stone_2026,Zhu:2024utz, Fields:2024pob} merger simulation, recovers the remnant structure, neutrino luminosities, and ejecta composition of a reference run using the full \nurates library.

This paper is organized as follows. In Sec.~\ref{sec:methods} we describe the \nurates opacity library and the gray reaction rates it provides, and introduce our neural network emulator, detailing its architecture and the construction of the training data. In Sec.~\ref{sec:results} we benchmark the emulator's computational performance and assess its accuracy against the full library on an independent simulation snapshot. We further demonstrate its use in a global BNS merger simulation by comparing the remnant properties, neutrino luminosities, and ejecta properties against a reference run.  We summarize our findings and discuss future work in
Sec.~\ref{sec:conclusion}. Training and hyperparameter tuning are discussed in Appendices \ref{sec:training} and \ref{sec:ablation}.

\begin{figure*}
\includegraphics[width=2.0\columnwidth]{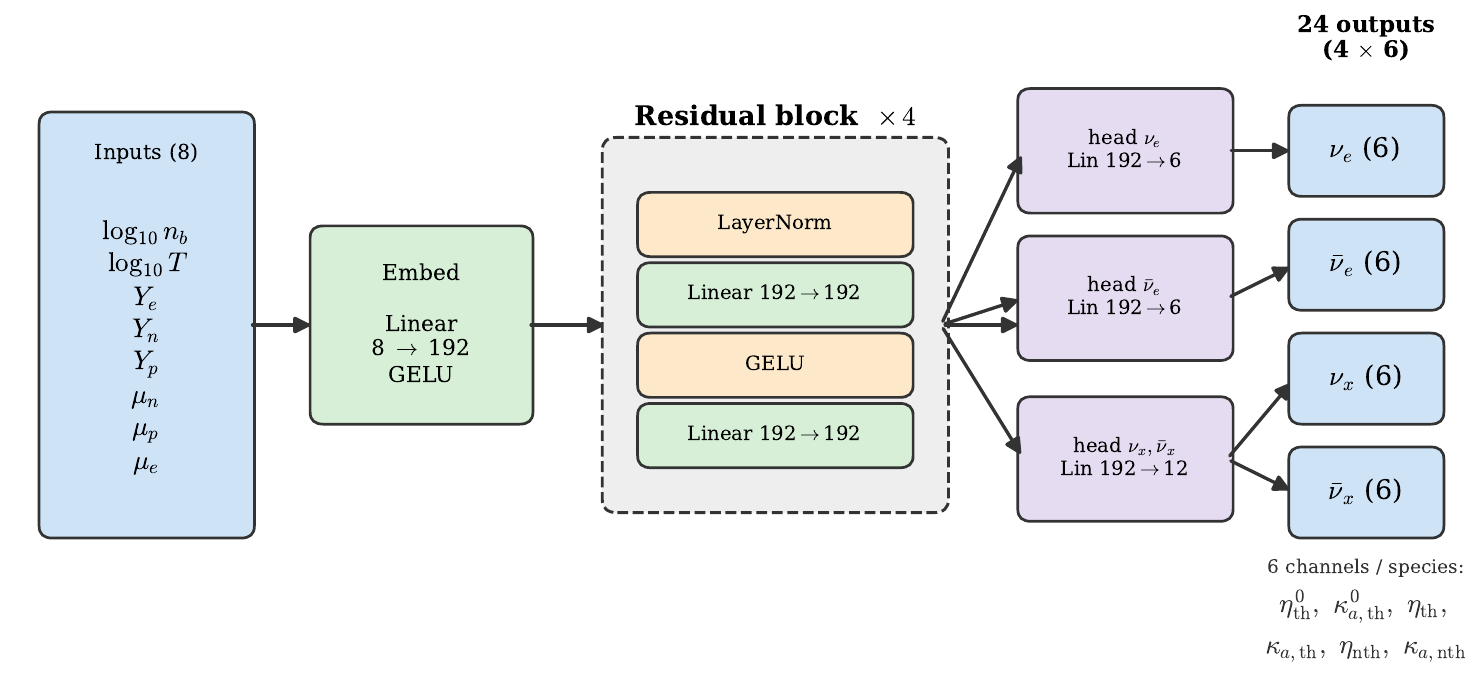}
\caption{Schematic of the emulator. The $8$ thermodynamic inputs are embedded into a hidden layer of width $h=192$ and passed through $N_{\rm blocks}=4$ residual blocks. Three linear heads ($\nu_e$, $\bar\nu_e$, and a shared $\nu_x,\bar\nu_x$ head) then produce $24$ outputs: $6$ channels for each of the $4$ species, namely the thermal $\eta^0_{\rm th}$, $\kappa^0_{a,\rm th}$, $\eta_{\rm th}$, $\kappa_{a,\rm th}$ and the non-thermal $\eta_{\rm nth}$, $\kappa_{a,\rm nth}$. All channels are predicted independently.}
\label{fig:arch}
\end{figure*}

\section{Methods} \label{sec:methods}

\subsection{\nurates gray opacity}

\nurates~\cite{Chiesa:2024lnu} is an open-source, performance-portable~\cite{KokkosCore2022} library for computing the neutrino interaction rates relevant to BNS mergers. It provides both energy-dependent (spectral) and energy-integrated (gray) emissivities $\eta$, absorption opacities $\kappa_{a}$ and scattering opacities $\kappa_s$. For energy-dependent rates, detailed balance relates the emission and absorption kernels, as well as the in- and out-scattering kernels, for each relevant process.

For a gray M1 neutrino transport scheme, the source term can be expressed as:
\begin{equation}
S^{\alpha} = \left(\eta - \kappa_{a}\, J\right) u^\alpha - \left(\kappa_{a} + \kappa_{s}\right) H^{\alpha} \, ,
\end{equation}
where $J$ and $H^\alpha$ are the neutrino energy density and flux in the fluid frame, and $u^\alpha$ is the fluid four-velocity. Furthermore, gray transport schemes usually account for the evolution of the neutrino number density as well~\cite{Foucart:2016rxm,Radice:2021jtw}. In this case, the source term is
\begin{equation}
\tilde{S} = \eta^{0} - \kappa^{0}_{a}\, n \, ,
\end{equation}
where $n$ is the neutrino number density in the fluid frame. Gray number rates are also computed using \nurates and are denoted here with a superscript $0$.

In this work, we consider a framework that features 4 distinct neutrino species: $\nue, \anue, \nux$ and $\anux$, where $\nux$ is a collective species for heavy-lepton neutrinos (muon and tau neutrinos combined) and $\anux$ for heavy-lepton anti-neutrinos.
The reactions that we consider are summarized in Table.~\ref{tab:reactions}. The charged-current beta processes and the iso-energetic (quasi-elastic) scattering of neutrinos on nucleons only involve a single integral over the neutrino energy each and are therefore inexpensive 1D kernels. By contrast, the pair processes, nucleon-nucleon bremsstrahlung and NEPS kernels are 2D, since each of them requires a double integral over the two neutrino phase spaces. 2D integrals are evaluated by nested Gauss-Legendre quadrature. To achieve better accuracy, the integration region is divided into two intervals around the maximum value of the integrand, with $N_q$\footnote{$N_q=6$ is set by default, with which one can achieve an accuracy of the order of $1\%$~\cite{Chiesa:2024lnu}.} nodes used in each interval. Consequently, the quadrature is performed using $2 \cdot N_q$ nodes per dimension and computational cost scales as $\mathcal{O}(N_q^2)$ kernel evaluations per cell, per process, and per neutrino species, against $\mathcal{O}(N_q)$ for the 1D kernels. Notably, our per-cell timing of the full library at $N_q=6$ shows that they account for more than $96\%$ of the total opacity wall time\footnote{Note that all three 2D processes require $\mathcal{O}(N_q^2)$ kernel evaluations, and the timing differences in Table.~\ref{tab:reactions} arise from the internal structure of the kernels rather than from the scaling~\cite{Chiesa:2024lnu}.}, while the 1D $\beta$ and iso-energetic scattering kernels together take less than $4\%$, as shown in Table.~\ref{tab:reactions}. This strongly motivates our hybrid strategy: we emulate with the NN only the three expensive 2D processes, whereas the cheap 1D kernels are still evaluated directly with \nurates.

\nurates groups the 2D processes into a thermal (th) and a non-thermal (nth) part:
\begin{equation}
\begin{aligned}
\eta &= \eta_{\rm th} + \eta_{\rm nth}\,, &
\kappa_a &= \kappa_{a,\rm th} + \kappa_{a,\rm nth}\,, \\
\eta^{0} &= \eta^{0}_{\rm th} + \eta^{0}_{\rm nth}\,, &
\kappa^{0}_a &= \kappa^{0}_{a,\rm th} + \kappa^{0}_{a,\rm nth}\,.
\end{aligned}
\end{equation}
The thermal part collects pair processes and bremsstrahlung. These reactions are mostly relevant in regions where neutrino and antineutrino pairs are created or annihilated by the thermal bath of $e^\pm$ or nucleons. When the medium is in thermal equilibrium with the neutrino radiation, Kirchhoff's law can be applied to directly express the opacity in terms of the emissivity, and vice versa~\cite{Radice:2021jtw}. The non-thermal part contains the NEPS, a process mostly relevant in scattering atmospheres in which matter and neutrino radiation are not in thermal equilibrium.
Moreover, their contributions to the gray emissivity and opacity originate from in- and out-scattering kernels. Therefore they must be understood as effective absorption and emission gray rates and, within this approximate approach, they are not tied together by Kirchhoff's law. 
Overall, in the \nurates outputs, each species carries six gray channels, namely
$\{\eta^{0}_{\rm th},\kappa^{0}_{a,{\rm th}},\eta_{\rm th},\kappa_{a,{\rm th}},\eta_{\rm nth},\kappa_{a,{\rm nth}}\}$. Note that we exclude $\{\eta^{0}_{\rm nth},\kappa^{0}_{a,{\rm nth}}\}$ for the non-thermal channels since NEPS conserves the neutrino numbers, so our M1 scheme does not use them in the evolution of the neutrino number density. The total 2D output is therefore $4~\mathrm{species} \times 6~\mathrm{channels} = 24$ values per cell. The beta processes and iso-energetic scattering are added separately outside the $24$ emulated channels.

In general, the evaluation of the gray opacities and emissivities require the knowledge of the neutrino distribution function in the initial and final state. \nurates enables the computation of gray rates via two approaches: by assuming neutrinos to be in equilibrium with the fluid, or by reconstructing the neutrino distribution function from the gray radiation moments. The reconstruction method does not assume local thermodynamic equilibrium (LTE) between matter and radiation, thus it can be used without introducing ad hoc corrections to the LTE rates. However, because it requires more inputs from neutrino distributions, in what follows, we focus on the former approach. When computing rates assuming the neutrinos to follow the equilibrium distribution set by the local thermodynamic conditions, \nurates takes 8 parameters from the cell: the baryon number density $n_b$, temperature $T$, electron fraction $\Ye$, neutron and proton fractions $Y_n$ and $Y_p$, and the neutron, proton, and electron chemical potentials $\mu_n$, $\mu_p$, $\mu_e$. Note that only $n_b$, $T$ and $\Ye$ are independent. Since we assume nuclear statistical equilibrium (NSE), the composition and the chemical potentials are uniquely determined by $(n_b, T, \Ye)$.

\begin{table*}[htbp]
\centering
\caption{Reaction set in \nurates~\cite{Chiesa:2024lnu}, with the integral dimensionality and relative cost as tested for one zone on one GPU. Note that here $N$ stands for nucleon, and $\nu$ without subscript refers to any flavor of neutrinos.}
\label{tab:reactions}
\begin{tabular*}{\textwidth}{l@{\extracolsep{\fill}}lcc}
\hline \hline
Class & Reaction & Integration & Fraction of opacity kernel \\
\hline
Beta processes & $\begin{array}{ll}
\mathrm{p}+\mathrm{e}^{-} \leftrightarrow \mathrm{n}+\nu_{\mathrm{e}} & \mathrm{n}+\mathrm{e}^{+} \leftrightarrow \mathrm{p}+\bar{\nu}_{\mathrm{e}} \\
\mathrm{p} \leftrightarrow \mathrm{e}^{+}+\mathrm{n}+\nu_{\mathrm{e}} & \mathrm{n} \leftrightarrow \mathrm{e}^{-}+\mathrm{p}+\bar{\nu}_{\mathrm{e}}
\end{array}$ & 1D & $\sim1.3\%$ \\
Iso-energetic scattering & $\nu + N \rightarrow \nu + N$ & 1D & $\sim2.0\%$ \\
Pair processes & $\mathrm{e}^{-} + \mathrm{e}^{+} \leftrightarrow \nu + \bar{\nu}$ & 2D & $\sim17\%$ \\
Bremsstrahlung & $N + N \leftrightarrow N + N + \nu + \bar{\nu}$ & 2D & $\sim47\%$ \\
NEPS & $\mathrm{e}^{\pm} + \nu \leftrightarrow \mathrm{e}^{\pm} + \nu$ & 2D & $\sim33\%$ \\
\hline \hline
\end{tabular*}
\end{table*}

\subsection{Neural emulator}

Our emulator is a residual multilayer perceptron (MLP), shown schematically in Fig.~\ref{fig:arch}. Since $n_b$ and $T$ span several orders of magnitude, we use their logarithms as inputs, while the remaining inputs are used directly. Each input $x_i$ is then standardized as $\hat{x}_i = (x_i - \bar{x}_i)/s_i$, where $\bar{x}_i$ and $s_i$ are the mean and standard deviation of the $i$-th input feature over the training set~\cite{LeCun:1998}. This rescaling brings all inputs to zero mean and unit variance, which improves the conditioning of the optimization. The network consists of three stages. An embedding layer first maps the $8$ standardized inputs to a hidden layer of width $h=192$, followed by a Gaussian error linear unit (GELU) activation~\cite{Hendrycks:2016gelu}. The hidden state is then processed by $N_{\rm blocks}=4$ residual blocks~\cite{He:2016resnet}, each of which applies $\mathrm{LayerNorm}(h)\to\mathrm{Linear}(h,h)\to\mathrm{GELU}\to\mathrm{Linear}(h,h)$, with LayerNorm denoting layer normalization~\cite{Ba:2016layernorm}, and adds the result back to its input. Finally, the shared hidden state is passed to three separate linear output heads: one for $\nu_e$ ($h\to6$), one for $\bar\nu_e$ ($h\to6$), and one shared by the heavy-lepton species $\nu_x$ and $\bar\nu_x$ ($h\to12$). Separate heads give each species group its own output projection at negligible extra cost. Together they produce $24$ outputs, i.e.\ $6$ channels for each of the $4$ species. All $24$ channels, including the $\nu_x$ and $\bar\nu_x$ thermal channels, are predicted independently, without imposing $\nu_x$--$\bar\nu_x$ symmetry. The network outputs (standardized) base-10 logarithms of the rates rather than the rates themselves. Since the rates vary by many orders of magnitude across the thermodynamic conditions of BNS mergers, predicting their logarithms compresses the dynamic range the network has to learn and makes the training loss sensitive to relative rather than absolute errors. In total, the network has ${\sim}304$K trainable parameters. Note that this architecture is not unique. To justify our design choices, we present our ablation study in Appendix.~\ref{sec:ablation}.

We generate the opacities used as training data starting from an ample set of thermodynamic points $(n_b, T, \Ye)$ and inputting them to \nurates. The training sets therefore inherit the accuracy of \nurates at the chosen quadrature order, for which we adopt $N_q=$ 6 or 30 depending on the analysis, as discussed later. The chemical potentials and nucleon fractions are obtained from the finite temperature, composition-dependent DD2 equation of state~\cite{Typel:2009sy, Hempel:2009mc} through the CompOSE/PyCompOSE interface~\cite{CompOSE2022}, given $(n_b, T, \Ye)$. Consequently, the emulator is tied to the DD2 equation of state table it was trained on. The loss function is a weighted (MSE) in logarithmic (Appendix.~\ref{sec:training}).

The set of relevant thermodynamic conditions is collected from two sources: points taken from BNS merger simulation snapshots\footnote{These snapshots are taken from an independent global BNS merger simulation, which adopts the same setup as in Sec.~\ref{sec:insim} but an earlier version of \nurates. Note that this difference affects, if any, only the sampled thermodynamic conditions and not the model fidelity, since all rates are recomputed with the current version of \nurates.}, and points drawn from specific grids in the $(n_b, T, \Ye)$ space. To sample the full geometry of a BNS merger remnant, we read primitive variable snapshots (density, temperature, electron fraction) in both the equatorial ($xy$) plane, which captures the disk and equatorial outflows, and the meridional ($xz$) plane, which captures the polar winds with high-$\Ye$ matter. Snapshots are taken at 39 times spanning $t \approx 1$--$39$~ms post-merger. The baseline training set consists of data from $985{,}561$ equatorial cells and $61{,}155$ meridional cells. Additionally, because the early post-merger snapshots under-represent the hottest matter, we extract $285{,}389$ cells restricted to the hot core ($T \geq 10$~MeV) across the evolution. In total the simulation contributes $1{,}332{,}105$ cells ($96.9\%$ of the training set). Two regular thermodynamic grids supplement the snapshots in regimes that remain sparsely sampled. A low-temperature grid of $12{,}000$ points tiles $T \in [0.10, 0.20]$~MeV over $\log_{10} n_b \in [-12, -0.5]~\mathrm{fm^{-3}}$ and $\Ye \in [0.05, 0.55]$, filling gaps in the cold ejecta. A high-density grid of $30{,}000$ points covers the dense remnant regime $n_b \in [0.3, 1.0]~\mathrm{fm^{-3}}$, $T \in [1, 100]$~MeV, $\Ye \in [0.01, 0.60]$. Together, these grids contribute $42{,}000$ points (${\sim}3.1\%$ of the combined set).

The simulation and grid data are combined into a dataset of $1{,}374{,}105$ labeled points. A fixed random split assigns $80\%$ ($1{,}099{,}284$ points) to training, $10\%$ ($137{,}410$) to validation (used for early stopping and hyperparameter selection), and $10\%$ ($137{,}411$) to an internal test set. Because the training and test cells are drawn from the same snapshots, we additionally assess final accuracy on an \emph{independent} simulation snapshot at merger time ($48{,}683$ equatorial cells) that is excluded from training entirely, providing a fair benchmark. We also provide more training details in Appendix.~\ref{sec:training}.

Once trained, the network is serialized to a portable ${\sim}1.3$~MB size TorchScript~\cite{Paszke:2019xhz} to be integrated in global simulations. Being self-contained, the serialized module is loaded and evaluated at runtime without any Python dependency. The emulator is embedded in the gray M1 neutrino transport module of \athenak~\cite{Stone_2026,Zhu:2024utz, Fields:2024pob}, which is a GPU-portable general relativistic magnetohydrodynamics code. The M1 scheme is ported from and broadly consistent with \texttt{THC\_M1}~\cite{Radice:2021jtw}. Inference is performed on the GPU: at each cycle the active cells belonging to a given rank are gathered and passed through the network in a single batched evaluation, so that the emulator can be ran collectively.

\begin{figure*}
\includegraphics[width=1.5\columnwidth]{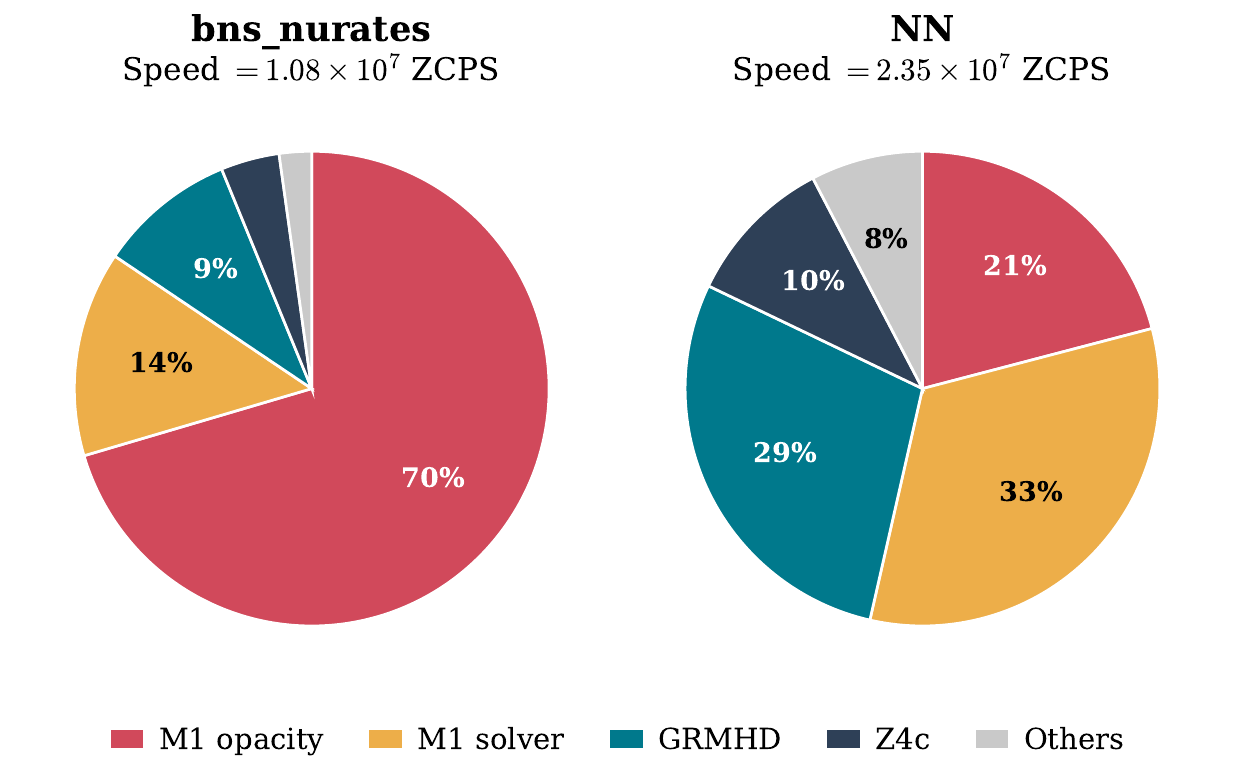}
\caption{Simulation speed in zone-cycles per second and breakdown of Kokkos kernel time over $20$ cycles of a BNS run on $16$ A100 GPUs, for the \nurates baseline (left) and the hybrid NN model (right). Kernels are grouped into M1 opacity, M1 solver, GRMHD, Z4c, and Others (see text for details). In the baseline, the opacity calculation takes $70\%$ of the kernel time. Remarkably, with the NN, it falls to ${\approx}21\%$ (NN inference plus the exact 1D processes). The opacity calculation using NN forward pass alone is ${\approx}12$ times cheaper than the 2D \nurates kernel it replaces, giving a ${\approx}2.2\times$ overall speedup of the simulation. }
\label{fig:timing}
\end{figure*}

\section{Results} \label{sec:results}

\begin{table*}[htbp]
\centering
\caption{Emulator accuracy on the independent snapshot at $t - t_\mathrm{merger} \approx 0.5$~ms ($163{,}410$ cells), given as median/P90 relative error [\%]. Columns are the number ($\eta^{0}$, $\kappa^{0}_a$) and energy ($\eta$, $\kappa_a$) emissivities and absorption opacities, split into thermal (th) and non-thermal (nth) parts. The non-thermal number channels are not emulated (Sec.~\ref{sec:methods}). Upper panel: error of the combined rates (1D processes from \nurates plus emulated 2D integrals trained with $\Nq=30$) with respect to \nurates at $\Nq=30$. Lower panel: quadrature error of \nurates, i.e., the combined rates at $\Nq=6$ with respect to $\Nq=30$, shown as a reference.}
\label{tab:nnacc}
\begin{tabular*}{\textwidth}{l@{\extracolsep{\fill}}cccccc}
\hline \hline
Species & $\eta^{0}_{\rm th}$ & $\kappa^{0}_{a,\rm th}$ & $\eta_{\rm th}$ & $\kappa_{a,\rm th}$ & $\eta_{\rm nth}$ & $\kappa_{a,\rm nth}$ \\
\hline
\multicolumn{7}{l}{\textit{NN emulator vs \nurates, both with $\Nq=30$ — median/P90 [\%]}} \\
$\nue$  & 0.08/12.8 & 0.09/11.2 & 0.05/12.6 & 0.06/11.2 & 3.16/20.7 & 1.35/6.3 \\
$\anue$ & 0.14/7.3  & 0.17/8.4  & 0.10/7.1  & 0.14/8.9  & 2.88/14.2 & 1.84/8.0 \\
$\nux$  & 1.87/7.1  & 1.51/6.5  & 2.09/7.1  & 1.51/6.6  & 1.58/6.5  & 1.55/6.8 \\
$\anux$ & 1.87/7.1  & 1.51/6.5  & 2.08/7.1  & 1.50/6.6  & 1.49/6.8  & 1.53/6.8 \\
\hline
\multicolumn{7}{l}{\textit{Quadrature error for \nurates, with $\Nq=6$ vs $\Nq=30$ — median/P90 [\%]}} \\
$\nue$  & 0.09/59.0 & 0.08/53.8 & 0.31/56.4 & 0.31/53.3 & 0.12/0.8  & 0.12/0.8 \\
$\anue$ & 0.12/2.9  & 0.12/0.8  & 0.39/3.0  & 0.38/2.2  & 0.20/21.5 & 0.20/21.5 \\
$\nux$  & 0.29/0.5  & 0.29/0.5  & 2.21/2.6  & 2.21/2.6  & 0.12/0.2  & 0.12/0.2 \\
$\anux$ & 0.29/0.5  & 0.29/0.5  & 2.21/2.6  & 2.21/2.6  & 0.13/0.2  & 0.13/0.2 \\
\hline \hline
\end{tabular*}
\end{table*}

\begin{figure*}
\includegraphics[width=2.0\columnwidth]{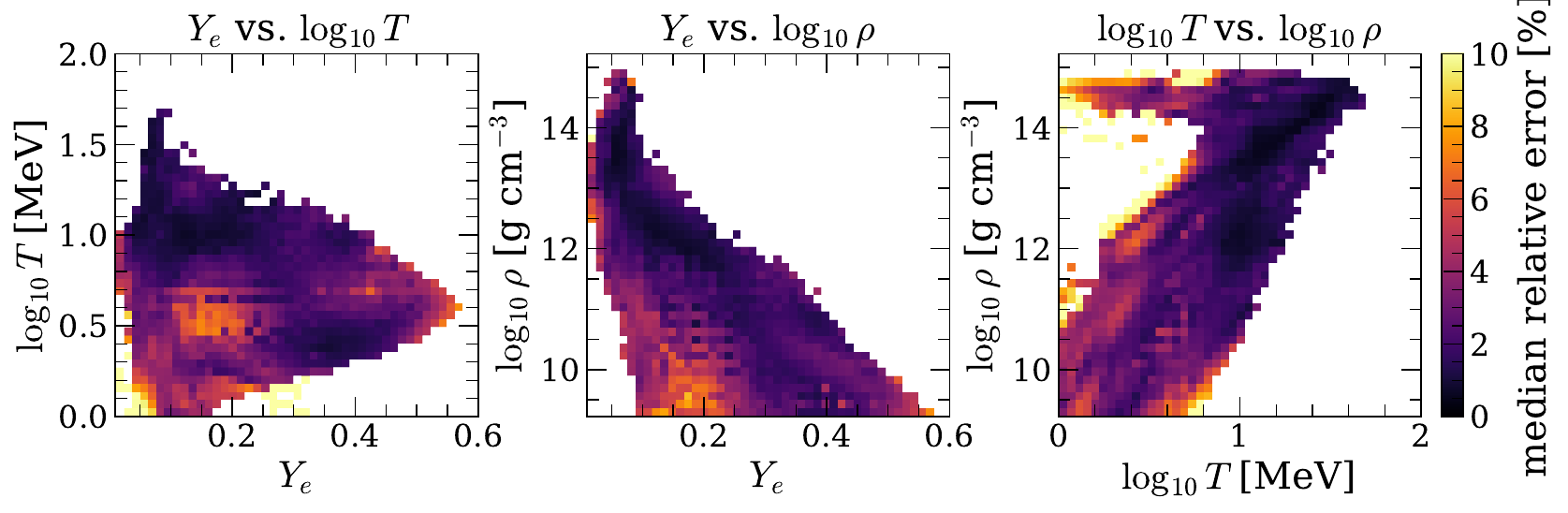}
\caption{Median relative error of the combined (emulated 2D $+$ \nurates 1D) rates, averaged over all $24$ channels, across the thermodynamic states reached in a BNS simulation, projected onto the $\Ye$--$T$ (left), $\Ye$--$\rho$ (middle), and $T$--$\rho$ (right) planes, where $\rho$ is the rest-mass density in CGS units, to make more physical sense. 
}
\label{fig:errmap}
\end{figure*}

\subsection{GPU performance benchmarks}

We compare the production performance of the full \nurates calculation and the hybrid NN model in a 3D BNS merger simulation on a domain spanning $[-1024,1024]M_\odot$ per dimension, with $7$ levels of adaptive mesh refinement (AMR)~\cite{Berger:1984zza,Berger:1989a} and otherwise the same setup as the production merger simulation that will be discussed in Sec.~\ref{sec:insim}. Timings are collected using KokkosTools~\cite{KokkosTools2018} over $20$ cycles on $16$ A100 GPUs, excluding the first $5$ cycles to avoid initialization overheads, and are shown in Fig.~\ref{fig:timing}. For clarity, we group the Kokkos kernels into five categories. \emph{M1 opacity} contains the computation of the neutrino emissivities and opacities, using either \nurates (left) or our hybrid NN (right). \emph{M1 solver} contains the remaining M1 radiation transport: the implicit update of the radiation moments, the closure solve, the radiation fluxes, and the radiation stress-energy coupling. \emph{GRMHD} contains the evolution of the general relativistic magnetohydrodynamics equations~\cite{Fields:2024pob} including the conservative-to-primitive inversion and the update of the matter and magnetic field. \emph{Z4c} contains the spacetime evolution and constraint evaluation~\cite{Zhu:2024utz}. \emph{Others} contains other kernels like AMR prolongation/restriction.

In the baseline run, the \nurates opacity kernel takes $532.0$~s, or $70.4\%$ of the total kernel time of $755.3$~s, making it by far the dominant cost. In the hybrid run, the whole hybrid M1 opacity step takes $61.3$~s of $293.0$~s ($20.9\%$). This consists of $43.9$~s of NN inference, $15.6$~s for the exactly evaluated 1D processes, and $1.8$~s of input packing and output readout. The opacity calculation is therefore ${\approx}8.7$ times cheaper overall, and the NN forward pass alone is ${\approx}12$ times cheaper than the 2D \nurates kernel it replaces. As a result, the M1 opacity is no longer the bottleneck: it now ranks behind both the M1 solver ($95.6$~s, $32.6\%$) and GRMHD ($83.7$~s, $28.6\%$). The total kernel time drops from $755.3$~s to $293.0$~s, and the simulation speed increases from $1.085\times10^{7}$ to $2.347\times10^{7}$ zone-cycles per second (ZCPS), giving a measured end-to-end speedup of $2.16$.

\begin{figure*}
\includegraphics[width=2.0\columnwidth]{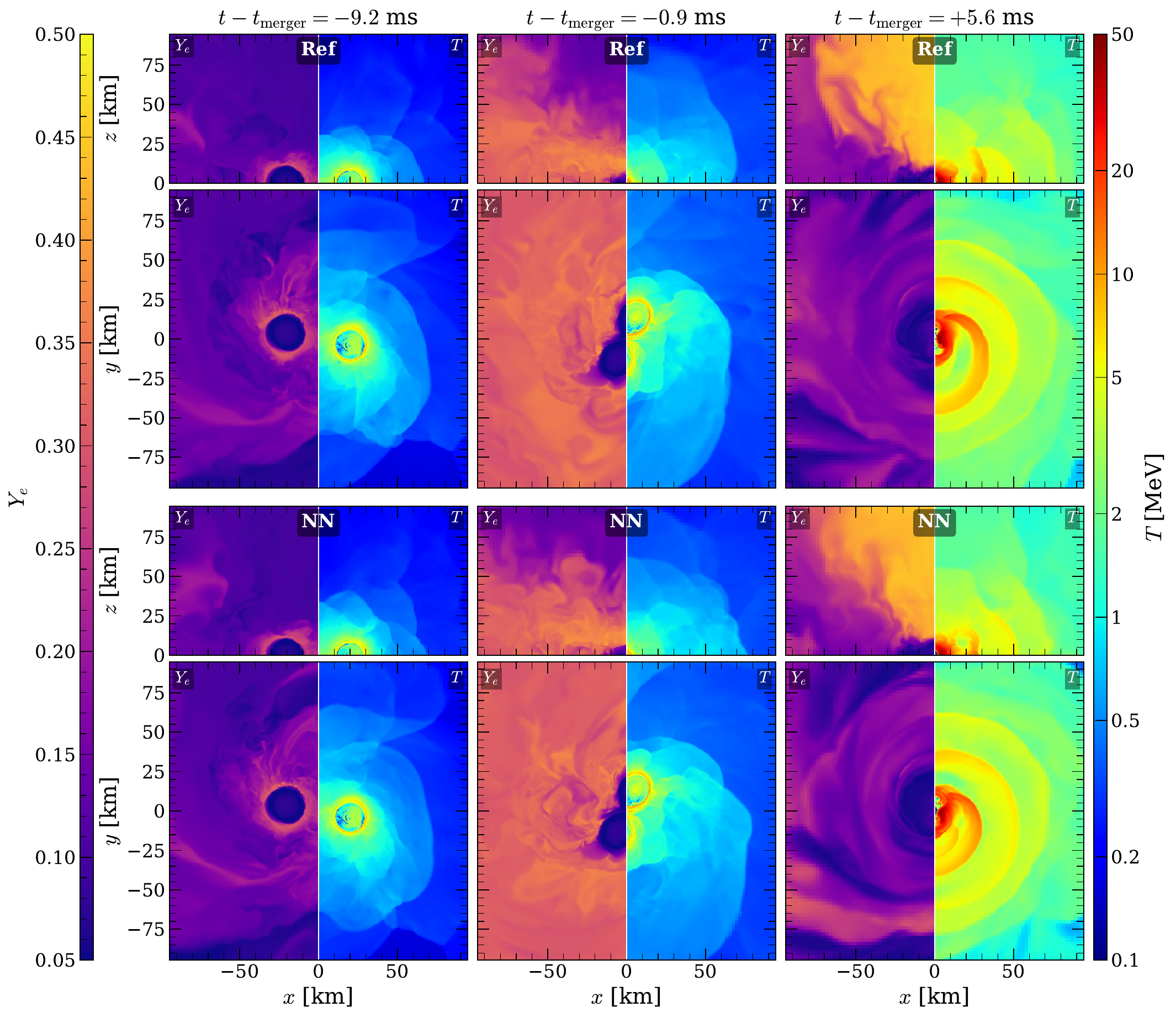}
\caption{Electron fraction (left half of each panel) and temperature (right half of each panel) of the reference (upper rows) and NN (lower rows) runs, $9.2$~ms and $0.9$~ms before merger and $5.6$~ms after merger (columns). Each run is shown in the meridional ($xz$, $z \geq 0$) and equatorial ($xy$) planes in each panel's upper half and lower half parts, respectively. We see, in general, qualitatively the same $\Ye$ and temperature distributions throughout the inspiral, merger and post-merger phases.}
\label{fig:insim_snap}
\end{figure*}

\subsection{Neural network accuracy}

We evaluate the hybrid emulator trained with $\Nq=30$ on an independent equatorial snapshot at $t - t_\mathrm{merger} \approx 0.5$~ms (containing $163{,}410$ cells), which is excluded from training. Since it is taken shortly after merger, this snapshot not only represents the most physically crucial phase, but also covers a broad range of densities, temperatures, and electron fractions. To quantify the accuracy of the emulator, we compute, for each cell $i$, neutrino species, and channel, the relative error $\delta_i = |X_i^{\rm NN} - X_i^{\rm ref}|/X_i^{\rm ref}$, where $X$ denotes any of the gray rates, $X^{\rm NN}$ is the combined value (1D processes from \nurates plus emulated 2D integrals), and $X^{\rm ref}$ is the full \nurates evaluation at $\Nq=30$. Table~\ref{tab:nnacc} reports the median and the 90th percentile (P90) of the distribution of $\delta_i$ over all cells of the snapshot, i.e., the error below which $50\%$ and $90\%$ of the cells fall, respectively. While the median measures the typical accuracy, P90 is sensitive to the tail of the distribution, which is dominated by the regions where the emulator performs worst. 

In the electron flavor thermal channels, which are dominated by the $\beta$ processes computed with \nurates, the hybrid method's error is diluted to the sub-percent level, with median errors $\lesssim 0.2\%$ for $\nue$ and $\anue$. The P90 of these channels is nevertheless $7$--$13\%$, which suggests that a non-negligible fraction of cells lies in regions where the emulated pair and bremsstrahlung contributions are comparable to those of the $\beta$ processes, e.g., at low densities. The non-thermal channels of $\nue$ and $\anue$, which are fully emulated, show median errors of $1.4$--$3.2\%$ and P90 of $6$--$21\%$. Since heavy-lepton neutrinos do not undergo charged-current reactions, all their channels are set by the emulated 2D integrals, and they show median errors of $1.5$--$2.1\%$ and P90 of $6.5$--$7.1\%$ in both thermal and non-thermal channels. Note that the thermal channels of $\nux$ and $\anux$, which are predicted independently by the network, show nearly identical error statistics. The largest median ($3.2\%$) and P90 ($20.7\%$) are both found for the non-thermal energy emissivity of $\nue$. 

To put these numbers into context, we compare them with the quadrature error of \nurates itself, shown in the bottom panel of Table~\ref{tab:nnacc}. At $\Nq=6$, which was identified as a good trade-off between accuracy and cost for gray M1 schemes~\cite{Chiesa:2024lnu}, the median quadrature error is below $0.4\%$ in all channels except the heavy-lepton thermal energy channels, where it reaches $2.2\%$. The P90, however, ranges from $\lesssim 3\%$ for heavy-lepton neutrinos up to $53$--$59\%$ in the thermal channels of $\nue$ and $21.5\%$ in the non-thermal channels of $\anue$. The emulator error is therefore comparable to the quadrature error at $\Nq=6$ in most channels, larger by roughly one order of magnitude in the median of the non-thermal channels and smaller in the tails of the $\nue$ thermal and $\anue$ non-thermal channels, where the quadrature error is the largest. 

To identify where in the thermodynamic space the residual error lies, we further evaluate the emulator on a sample that covers the full extent of the states reached in the BNS merger. We collect the cells within $400\,\Msun$ (${\sim}600$~km) from the center from $66$ three-dimensional snapshots of a reference BNS simulation ranging from inspiral to post-merger phases ($t \simeq 0$--$57$~ms), which yields $81{,}897$ cells. We bin them on a $24^3$ grid in $(\log_{10}\rho, \log_{10}T, \Ye)$, and draw for each bin the averaged relative error of the combined (emulated 2D $+$ \nurates 1D) rates over all $24$ channels. 
In Fig.~\ref{fig:errmap}, we show the median averaged errors in three projections of the thermodynamic space. The median over all the cells is $2.4\%$, the P90 is $6.4\%$, and only $2\%$ of the cells have averaged error that exceeds $10\%$. In the hot and dense bulk of the remnant ($T \gtrsim 6$~MeV, $2\times10^{11} \lesssim \rho \lesssim 2\times10^{14}$~g~cm$^{-3}$), the error is ${\sim}1.3\%$, with a P90 below $3\%$. The error increases toward the edges of the matter distribution, where the rates vary the most, e.g., looking at the rightmost panel, where $\rho \gtrsim 2\times10^{12}$~g~cm$^{-3}$ and at $T \lesssim 3$~MeV (median $5.6\%$), or looking at the leftmost panel, where $\Ye > 0.45$ (median $3.2\%$). There is also a feature that does not lie at an edge, which is a band at $\Ye \simeq 0.1$--$0.2$ and $T \simeq 3$--$5$~MeV ($\rho \simeq 3\times10^{9}$--$3\times10^{11}$~g~cm$^{-3}$), where the median error reaches ${\sim}6\%$, compared to ${\sim}2.4\%$ at the neighboring temperatures. This error is carried mostly by the heavy-lepton species ($7$--$8\%$), while the electron flavors are below $4\%$. The band is pronounced there, most likely because we up-weight the adjacent cells with $T < 3$~MeV and $\Ye < 0.1$ by a factor of $8$, as we will discuss in Appendix~\ref{sec:training}. Therefore, the errors in the regions around the band are much lower, leading to this distinct spot. Overall, the error of $\nue$ and $\anue$ remains at the few-percent level across the thermodynamic states reached in the simulation. Since the electron (anti-)neutrinos control the evolution of $\Ye$, we expect the uncertainties associated with the emulator to have a limited impact on the ejecta composition. We test this expectation directly in the global simulation in Sec.~\ref{sec:insim}.

\begin{figure}
\includegraphics[width=0.98\columnwidth]{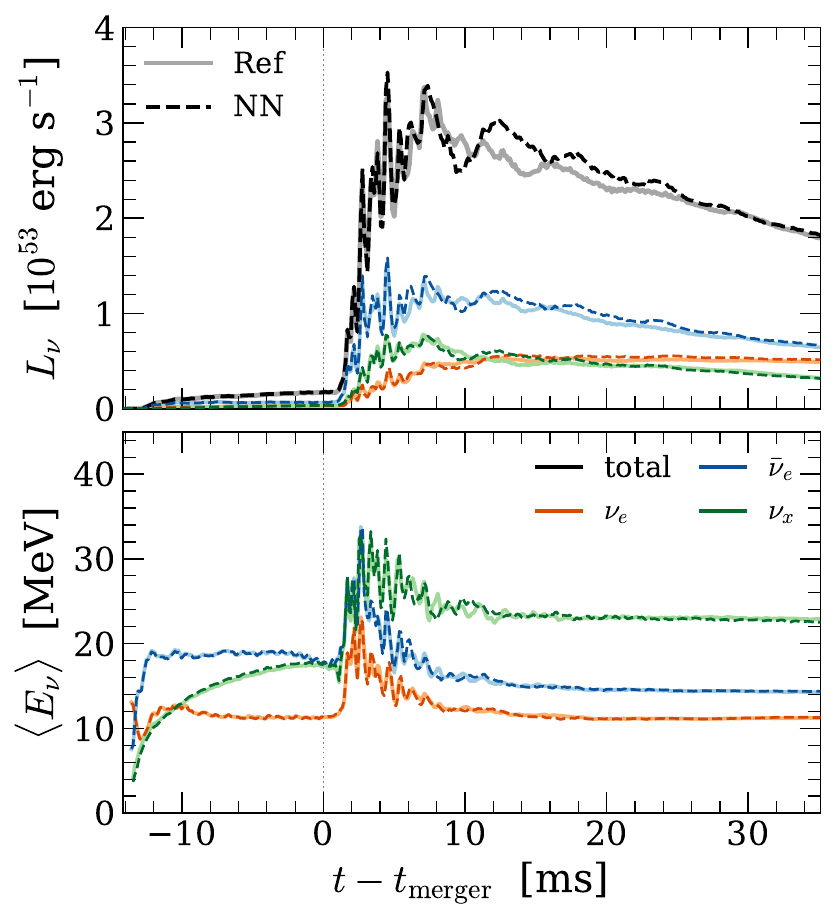}
\caption{Neutrino luminosities (upper panel) and mean energies (lower panel) versus time relative to merger, for $\nu_e$, $\bar{\nu}_e$, and $\nu_x\left(=\nu_\mu+\nu_\tau\right)$, extracted on a coordinate sphere of radius $200\,M_\odot \simeq 300$~km. The total luminosity in the upper panel is summed over all four evolved species (including $\bar{\nu}_x$). Light solid lines show the reference (Ref) run; dark dashed lines, drawn on top, show the NN run. The emulator reproduces the neutrino luminosities and mean energies, both in their hierarchy across species and in their qualitative trends.}
\label{fig:insim_lumi}
\end{figure}

\subsection{BNS merger simulation}
\label{sec:insim}

\begin{figure}
\includegraphics[width=0.98\columnwidth]{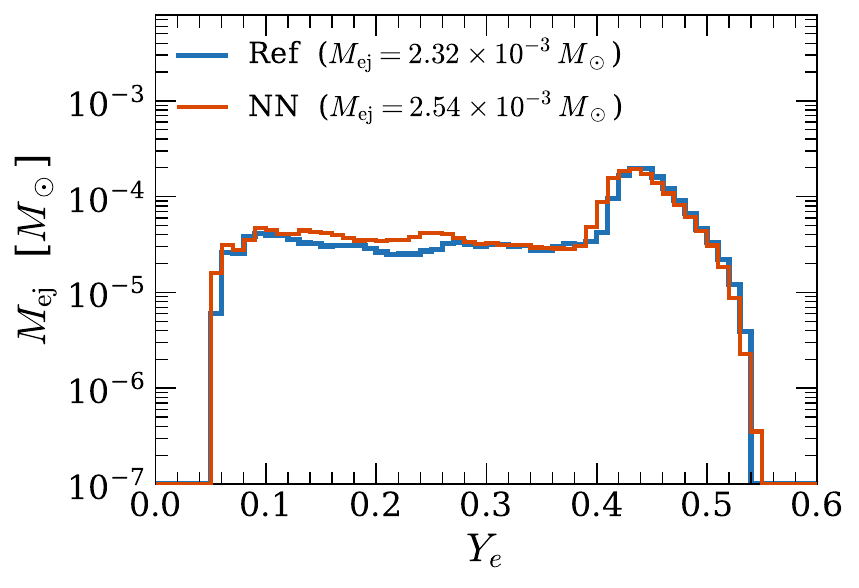}
\caption{Mass of the unbound ejecta per electron fraction bin ($\Delta Y_e = 0.01$) for the Ref and NN runs. Ejecta are measured with the geodesic criterion and collected on a coordinate sphere of radius $200\,M_\odot \simeq 300$~km from the start of the simulation until $25$~ms post-merger. The legend gives the total ejecta mass of each run.}
\label{fig:insim_ejecta}
\end{figure}

We present ab-initio global BNS merger runs that use our neural emulator trained with \nurates at $\Nq=6$ to replace \nurates in the opacity evaluation. We run with \athenak, evolving an equal mass $1.35\,M_\odot + 1.35\,M_\odot$, non-rotating binary with the DD2 equation of state. The stars are placed at an initial separation of $45\,\mathrm{km}$, with initial data generated using the \texttt{Lorene} pseudo-spectral code~\cite{Gourgoulhon:2000nn}, and initial dipolar magnetic fields with $B_{\rm max}=10^{15} G$ (a similar post-amplification strength as the setup used in~\cite{Fields:2025uyk}). AMR is used with 9 refinement levels. The finest refinement region has grid spacings $0.167\,M_{\odot} \simeq 246\,\mathrm{m}$. 

The two stars merge ${\sim}14$~ms after the start of the simulation. We then evolve the remnants for an additional ${\sim}36$~ms to characterize the dynamical mass ejections and neutrino emission~\cite{Radice:2016dwd,Radice:2018pdn,Sekiguchi:2016bjd}. The merger times of the reference and NN runs, measured from the peak of the $(2,2)$-mode GW strain amplitude, agree to within ${\simeq}0.01$~ms. Fig.~\ref{fig:insim_snap} compares the electron fraction and temperature of the two runs in the meridional ($xz$) and equatorial ($xy$) planes at three common coordinate times: early in the inspiral ($9.2$~ms before merger), just before contact ($0.9$~ms before merger), and $5.6$~ms after merger. During the inspiral the two runs are nearly indistinguishable. Inside the stars, the $Y_e$ differences are below $0.01$ everywhere and the temperature differences are below $0.5$~MeV for most of the area. After merger, the simulation employing the emulator recovers the large-scale structure of the remnant, the forming disk with spiral arms, and the polar neutrino-driven outflow. In the disk ($10^{11} < \rho < 10^{13}\,\mathrm{g\,cm^{-3}}$), the median differences are $|\Delta T| \simeq 0.5$~MeV and $|\Delta Y_e| \simeq 0.02$--$0.03$. The $Y_e$ of the hot, optically thick remnant also agrees to median $|\Delta Y_e| \simeq 0.005$. There are cells in the remnant where temperature shows larger local differences, because the hot spots of the oscillating remnant sit at slightly different positions in the two runs. These differences depend on the phase of the turbulent post-merger dynamics. However, as we show below, these unavoidable local deviations have only a limited impact on the integrated neutrino emission and on the composition of the ejecta.

Fig.~\ref{fig:insim_lumi} compares the neutrino luminosities of the two runs (the total, and separately for $\nue$, $\anue$, and $\nux$) together with their mean energies~\cite{Cusinato:2021zin}, extracted on a coordinate sphere of radius $200\,M_\odot \simeq 300$~km. The luminosities evolve in the same way in both runs. They are low during the inspiral, rise sharply within ${\sim}2$~ms after merger as the hot remnant forms, peak at a total of ${\simeq}3.4\times10^{53}\,\mathrm{erg\,s^{-1}}$, and then decline gradually. At $t - t_\mathrm{merger} > 2$~ms, the median relative difference between the two runs is $3.7\%$ for the total luminosity and $4.6$--$5.8\%$ for the individual species. The post-merger luminosities agrees to within $4.4\%$ for every species. The mean energies agree even more closely, with median relative differences of ${\lesssim}1\%$. The usage of the emulator thus preserves both the hierarchy of luminosities and mean energies across species with very high fidelity. 

Finally, Fig.~\ref{fig:insim_ejecta} compares the electron fraction distribution of the unbound ejecta in the two runs. We measure the outflow through a coordinate sphere of radius ${\simeq}300$~km and classify the matter as unbound using the geodesic criterion: $-u_t > 1$, where $u_t$ is the time component of the fluid four-velocity. The NN and reference run histograms have similar shapes, each with a low-$Y_e$ ($<0.2$) component from the dynamical ejecta and a high-$Y_e$ ($>0.4$) component from polar materials. The total ejected masses agree to within $10\%$, with $M^{\rm ej}_{\rm total}=2.32\times10^{-3}\,M_\odot$ for the reference run and $2.54\times10^{-3}\,M_\odot$ for NN. The composition agrees qualitatively as well. The mass fractions with $Y_e<0.2$ are $0.20$ (Ref) and $0.22$ (NN), and those with $Y_e>0.4$ are $0.54$ and $0.51$. The mass-weighted mean $Y_e$ differs by only $0.012$ ($0.348$ versus $0.336$), with the NN ejecta slightly more neutron rich. Since the electron fraction of the ejecta is the main driver of the $r$-process yields and of the kilonova light curves, this agreement indicates that the per-cell errors of the emulator do not accumulate into a significant systematic bias in the composition of the outflows. In particular, the differences between the two runs in the ejecta $\Ye$ are well below the $10$--$30\%$ uncertainties estimated from comparisons between M1 and Monte Carlo neutrino transport~\cite{Foucart:2024npn}. This supports the use of the emulator in production simulations aimed at predicting multi-messenger observables.

\section{Conclusion} \label{sec:conclusion}

We present the first neural-network emulator for the neutrino reaction rates that are used in M1 transport in binary neutron star merger simulations. By emulating only the computationally dominant 2D kernels---pair processes, bremsstrahlung, and inelastic neutrino-electron scattering---while evaluating the inexpensive 1D beta processes and scattering directly through \nurates, the hybrid scheme reproduces the full \nurates opacities to within a few percent across the bulk of the thermodynamic phase space and reduces the opacity evaluation cost by roughly an order of magnitude. Embedded in \athenak, the emulator cuts the opacity share of the runtime from ${\sim}70\%$ to ${\sim}20\%$ and delivers a factor-of-two speedup for the merger simulations.

In our featuring ab-initio merger simulations, the usage of the emulator recovers the large-scale structure of the remnant and disk, the neutrino luminosities and mean energies to within $5\%$, and the electron fraction distribution of the unbound ejecta. The per-cell opacity errors average out in the integrated emission rather than accumulating into a systematic bias, so the quantities that set the r-process yields and the kilonova signals will be preserved. The residual error concentrates in two physically extreme populations: the hottest matter near the remnant surface and cells of negligible absolute opacity, both of limited dynamical consequence.

Several extensions remain. The present emulator is tied to the DD2 equation of state and to the reaction set with which its training data were generated; broadening it to other equations of state, and to other \nurates kernels, including muonic interactions~\cite{Gieg:2024jxs,Pajkos:2024iry,Ng:2024zve,Gieg:2026beb} would widen its range of applicability. Suppressing the errors in the difficult low $Y_e$ regime through targeted sampling or physics-informed constraints, and testing a broader range of binary configurations, are natural next steps toward improving the NN opacities in production multi-messenger simulations.

\vspace{4em plus 3em minus 3em}
\begin{acknowledgements}
The authors acknowledge the use of Opus and Sonnet by Anthropic to accelerate
numerical implementation and data analysis. All scientific analyses, numerical
results, and conclusions were produced by the authors, who take full
responsibility for the contents of this work.
Y.Q.~acknowledges support from the
Department of Energy, Office of Science, Division of Nuclear Physics, under
Award Numbers DE-SC0024388 and DE-SC0026818. 
Y.Q.~also acknowledges the hospitality of the University of Trento and of the Trento Institute for Fundamental Physics and Applications (TIFPA-INFN), where part of this work was carried out, and support from the DOE-INFN Summer Exchange Program 2026 (INFN Call No.~28477).
D.R.~acknowledges support from NASA through Grant No. 80NSSC25K7213, from the
Department of Energy, Office of Science, Division of Nuclear Physics, under
Awards Numbers DE-SC0024388 and DE-SC0026818, and from the National Science Foundation under Grants
No. PHY‐2407681, PHY-2512802, PHY‐2608045, and PHY‐2621752.
Simulations were performed on Aurora at the Argonne Leadership Computing
Facility, which is a DOE Office of Science User Facility supported under
Contract DE-AC02-06CH11357. An award of computer time was provided by the INCITE
program. Simulations were also performed on Perlmutter, at the National Energy
Research Scientific Computing Center (NERSC), a Department of Energy User
Facility, using NERSC award ERCAP0031370.
\end{acknowledgements}

\appendix

\section{Training}
\label{sec:training}

To account for the many orders of magnitude spanned by the rates, we train the network by minimizing a weighted mean-squared error in logarithmic space,
\begin{equation}
\mathcal{L} = \frac{1}{N_{\rm cell}} \sum_{i} a_i \sum_{s,c} w_{s,c} \left(\hat{y}^{(i)}_{s,c} - y^{(i)}_{s,c}\right)^2,
\label{eq:loss}
\end{equation}
where $y^{(i)}_{s,c}$ is the base-10 logarithm of the 2D contribution to channel $c$ of species $s$ computed with \nurates, $\hat{y}^{(i)}_{s,c}$ is the corresponding network output, and the first sum runs over the $N_{\rm cell}$ cells of a batch. The channel weights $w_{s,c}$ normalize each channel by the variance of $y_{s,c}$ over the training set, and increase the weight of the electron (anti-)neutrino channels by a factor of $3$, since these channels set the evolution of $\Ye$ in the outflows. The cell weights $a_i$ increase the contribution of the cold, hot, and low-$\Ye$ cells, which are rare in the training set but dynamically relevant, by a factor of $8$. Since the loss is evaluated on logarithms, it measures relative rather than absolute errors.

\begin{figure}
\includegraphics[width=0.98\columnwidth]{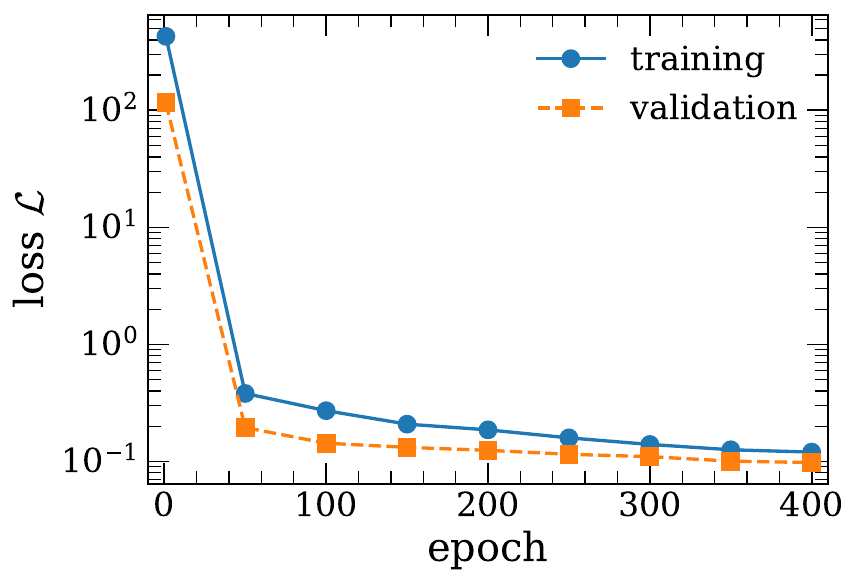}
\caption{Training and validation loss versus epoch for the production emulator recorded at the first epoch and every $50$ epochs thereafter. After a drop of about three orders of magnitude within the first $50$ epochs, both losses decrease slowly and show no late-time divergences until $400$ epochs. The validation loss lies slightly below the training loss because the training loss up-weights the rare cold, hot, and low-$\Ye$ cells.}
\label{fig:loss}
\end{figure}

We optimize the network with Adam~\cite{Kingma:2014adam}, using a weight decay of $10^{-4}$ and a batch size of $4096$. In Fig.~\ref{fig:loss}, we show the training and validation losses as a function of the epoch. The two curves track each other throughout training, and the validation loss does not increase at late times, which indicates that the network does not overfit. During $50$ to $400$ epochs, both curves asymptotically become flat, therefore we take $400$ as the stopping epoch. The weights are retained with the lowest validation loss. Note that the validation loss is computed without the cell weights $a_i$, and therefore lies slightly below the training loss, in which the hardest cells are up-weighted. 

\begin{table*}[htbp]
\centering
\caption{Ablation study. Each row varies one component relative to the production configuration ($8$ inputs, $h=192$, $N_{\rm blocks}=4$, $24$-channel head, trained with \nurates at $\Nq=30$). Columns give the median/P90 of the combined relative error $\mathcal{E}_s$ [\%] (eq.~\ref{eq:combined_err}) on the independent snapshot for each species, and the number of trainable parameters. Values are means over $3$ seeds, and the seed-to-seed spread of the median is below $1\%$ for all entries and typically ${\lesssim}0.6\%$.}
\label{tab:ablation}
\begin{tabular*}{\textwidth}{l@{\extracolsep{\fill}}ccccc}
\hline \hline
 & $\nue$ & $\anue$ & $\nux$ & $\anux$ & \# of parameters \\
Variant & med/P90 & med/P90 & med/P90 & med/P90 & (K) \\
\hline
\textbf{Production}                       & 0.7/2.0  & 0.8/1.9  & 2.5/5.4   & 2.3/5.2   & 304 \\
\hline
Table baseline (trilinear, $23^3$)        & 3.9/9.7  & 4.7/12.5 & 11.0/30.1 & 11.0/30.3 & --  \\
Emulate all channels (1D$+$2D)            & 2.1/4.4  & 2.5/5.3  & 2.8/5.6   & 2.8/5.5   & 304 \\
Target: linear$^\dagger$                  & 511/512  & 517/518  & 450/500   & 600/600   & 304 \\
Loss: no weights                          & 1.6/4.3  & 1.7/5.2  & 4.0/9.1   & 3.8/8.9   & 304 \\
Width $h=128$                             & 0.8/2.1  & 0.9/2.2  & 2.9/6.2   & 2.9/6.0   & 137 \\
Width $h=256$                             & 0.7/1.6  & 0.8/1.8  & 2.4/5.0   & 2.4/4.9   & 537 \\
Depth $N_{\rm blocks}=2$                  & 0.8/1.9  & 0.8/1.9  & 2.6/5.7   & 2.6/5.8   & 155 \\
Depth $N_{\rm blocks}=6$                  & 0.9/2.0  & 0.9/1.9  & 2.6/5.5   & 2.5/5.4   & 453 \\
No layer normalization                    & 0.7/1.9  & 1.1/2.3  & 2.4/5.8   & 2.5/5.7   & 303 \\
No residual connections                   & 0.8/1.9  & 1.0/2.0  & 2.4/5.3   & 2.3/4.9   & 304 \\
Data: $30\%$                              & 0.8/2.0  & 0.9/2.0  & 2.7/6.3   & 2.5/6.2   & 304 \\
Data: $10\%$                              & 1.0/2.4  & 1.3/2.8  & 4.2/9.1   & 4.0/9.0   & 304 \\
\hline \hline
\end{tabular*}
\end{table*}

\section{Ablation study}
\label{sec:ablation}

To justify the design of the emulator and to quantify the contribution of each of its ingredients, we retrain the network varying one component at a time, while keeping the optimizer, learning-rate schedule, training budget, early-stopping criterion, and data split fixed. Unless stated otherwise, the reference configuration is the production model described in Sec.~\ref{sec:methods}, i.e., $8$ inputs, hidden width $h=192$, $N_{\rm blocks}=4$ residual blocks, a $24$-channel output head without copied channels, and training labels computed at $\Nq=30$.

To ensure a fair comparison, we adopt the following protocol. First, every variant is evaluated on an independent equatorial snapshot at $t - t_\mathrm{merger} \approx 0.5$~ms ($163{,}410$ cells) of an BNS merger simulation that is not used for training, rather than on the random $80/10/10$ split. Second, each configuration is trained with $3$ random seeds. We report the mean over seeds. Third, we report accuracy and cost together. For accuracy, we use a single combined error per neutrino species. For each cell $i$ and species $s$, we average the relative error over the $N_{\rm ch}=6$ channels of that species,
\begin{equation}
\mathcal{E}^{(i)}_s = \frac{1}{N_{\rm ch}} \sum_{c=1}^{N_{\rm ch}} \frac{|X^{{\rm NN},(i)}_{s,c} - X^{{\rm ref},(i)}_{s,c}|}{X^{{\rm ref},(i)}_{s,c}}\,,
\label{eq:combined_err}
\end{equation}
where $X_{s,c}$ is the combined value of channel $c$ (emulated 2D integrals plus 1D processes from \nurates), i.e., the thermal number and energy emissivities and absorption opacities, and the non-thermal energy emissivity and absorption opacity, while the superscripts NN and ref denote the emulator and the full \nurates evaluation, respectively. We then report the median and the P90 of $\mathcal{E}_s$ over all cells for each of the four species. Since the $\beta$ processes are computed with \nurates, the median error of the electron-flavor species remains below $1\%$ for the production model, whereas the heavy-lepton species, which do not undergo charged-current reactions, directly reflect the accuracy of the emulated 2D integrals. And finally we also show the number of trainable parameters. All ablations are performed offline, and we do not repeat the merger simulation for the ablated variants, which would be computationally prohibitive.

Table.~\ref{tab:ablation} summarizes the study. To assess whether a neural network is needed at all, we compare it with direct trilinear interpolation in a 3D table. In the equilibrium mode adopted here, the rates depend only on $(n_b, T, \Ye)$, which makes tabulation the natural baseline. At a memory footprint equal to that of the network parameters (${\sim}1.2$~MB, i.e., a $23^3$ grid of single-precision values for each of the $24$ channels), the table is less accurate than the emulator by a factor of $4$--$6$ in the median (e.g., $11\%$ versus $2.3$--$2.5\%$ for the heavy-lepton species) and by a factor of $5$--$7$ in the P90. The largest errors of the table appear near the edges of the sampled thermodynamic range, possibly because the rates vary most steeply there. Note that the equilibrium mode is the most favorable case for tabulation. In the reconstruction mode, the rates also depend on the neutrino moments, so that the input dimension grows and tabulation quickly becomes impractical, which is the regime in which an emulator is expected to be much more advantageous.

To quantify the benefit of the hybrid strategy, we train a variant that emulates all channels, including the contributions of $\beta$ processes and iso-energetic scattering, on labels computed with all five reactions. Compared with the production model, this variant increases the median (P90) error of the electron species by a factor of ${\sim}3$ (${\sim}2.5$), i.e., from $0.7$--$0.8\%$ ($1.9$--$2.0\%$) to $2.1$--$2.5\%$ ($4.4$--$5.3\%$). The heavy-lepton species, whose emission and absorption receive no contribution from the 1D processes, are nearly unaffected, and the small increase of their median error (from $2.3$--$2.5\%$ to $2.8\%$) is possibly due to the network sharing its capacity among more targets. Since the 1D processes account for only ${\sim}3\%$ of the cost of the opacity evaluation (Table~\ref{tab:reactions}), computing them with \nurates comes at a negligible price.

To assess the choice of output space and loss weighting, we train two variants. The first predicts the rates in linear rather than logarithmic space, and fails: the network is wrong by orders of magnitude on all channels, exceeding $400\%$ median error. The second removes all weights from the loss, i.e., the per-channel variance normalization, the up-weighting of the electron-flavor channels, and the up-weighting of the hardest cells. This variant degrades more mildly, increasing the median (P90) error by a factor of $1.6$--$2.3$ ($1.7$--$2.7$), with the largest degradation found for the electron-flavor species, because their channels lose their additional weight. These results show that predicting $\log_{10}$ rates with a weighted loss is both essential to control the dynamic range of the rates, which span many orders of magnitude, and to guarantee the rate accuracies for the electron-flavor species.

To probe the dependence on network capacity, we vary the width ($h=128$, $256$) and the depth ($N_{\rm blocks}=2$, $6$) of the network. The error is comparable between $h=192$ and $h=256$ and between $4$ and $6$ blocks, and increases only slightly for the smaller networks, most visibly for the heavy-lepton species at $h=128$ ($2.9\%$ versus $2.3$--$2.5\%$ median). This saturation suggests that the accuracy of the emulator is not limited by its capacity and our current architecture is balanced. Removing either the layer normalization or the residual connections changes the median and P90 errors by less than $<0.1\%$. We nevertheless retain both, since they stabilize training and are expected to become important for deeper networks in our future work.

To quantify the role of the training data, we vary the amount of training data by reducing them randomly across the production training set. We find that training on $30\%$ of the data leaves the heavy-lepton median error nearly unchanged ($2.7\%$ versus $2.5\%$), while training on $10\%$ increases the mean error of all channels by $\geq 0.8\%$. Our production training data size therefore does not show sign of over-fitting. Together with the saturation of the capacity in our residual block, this suggests that the error in the bulk has nearly converged with respect to both the network size and the amount of data, and that the residual error is set mainly by the coverage of the rarest regions of the thermodynamic space. This motivates the targeted sampling that we plan to do in the future as discussed in Sec.~\ref{sec:conclusion}.

%%%%%%%%%%%%%%%%%%%%%%%%%%%%%%%%%%%%%%%%%%%%%%%%%%%%%%%%%%%%%%%%%%%%%
\bibliography{emulator.bib}

@article{MorawskiBejger2020,
  author  = {Morawski, F. and Bejger, M.},
  title   = {{Neural network reconstruction of the dense matter equation of state derived from the parameters of neutron stars}},
  journal = {Astron. Astrophys.}, volume = {642}, pages = {A78}, year = {2020},
  eprint  = {2006.07194}, archivePrefix = {arXiv}, doi = {10.1051/0004-6361/202038130}
}

@article{Radice:2016dwd,
    author = "Radice, David and Galeazzi, Filippo and Lippuner, Jonas and Roberts, Luke F. and Ott, Christian D. and Rezzolla, Luciano",
    title = "{Dynamical Mass Ejection from Binary Neutron Star Mergers}",
    eprint = "1601.02426",
    archivePrefix = "arXiv",
    primaryClass = "astro-ph.HE",
    doi = "10.1093/mnras/stw1227",
    journal = "Mon. Not. Roy. Astron. Soc.",
    volume = "460",
    number = "3",
    pages = "3255--3271",
    year = "2016"
}

@article{Foucart:2025nub,
    author = "Foucart, Francois",
    title = "{Assessing the difficulty of capturing the distribution function of neutrinos in neutron star merger simulations}",
    eprint = "2504.21822",
    archivePrefix = "arXiv",
    primaryClass = "astro-ph.HE",
    doi = "10.1103/gkk5-6wmh",
    journal = "Phys. Rev. D",
    volume = "112",
    number = "2",
    pages = "023046",
    year = "2025"
}

@article{Sekiguchi:2011zd,
    author = "Sekiguchi, Yuichiro and Kiuchi, Kenta and Kyutoku, Koutarou and Shibata, Masaru",
    title = "{Gravitational waves and neutrino emission from the merger of binary neutron stars}",
    eprint = "1105.2125",
    archivePrefix = "arXiv",
    primaryClass = "gr-qc",
    doi = "10.1103/PhysRevLett.107.051102",
    journal = "Phys. Rev. Lett.",
    volume = "107",
    pages = "051102",
    year = "2011"
}

@article{Rosswog:1998hy,
    author = "Rosswog, S. and Liebendoerfer, M. and Thielemann, F. K. and Davies, M. B. and Benz, W. and Piran, T.",
    title = "{Mass ejection in neutron star mergers}",
    eprint = "astro-ph/9811367",
    archivePrefix = "arXiv",
    journal = "Astron. Astrophys.",
    volume = "341",
    pages = "499--526",
    year = "1999"
}

@article{Ruffert:1998qg,
    author = "Ruffert, M. and Janka, H. T.",
    title = "{Gamma-ray bursts from accreting black holes in neutron star mergers}",
    eprint = "astro-ph/9809280",
    archivePrefix = "arXiv",
    journal = "Astron. Astrophys.",
    volume = "344",
    pages = "573--606",
    year = "1999"
}

@article{Shibata:2019wef,
    author = "Shibata, Masaru and Hotokezaka, Kenta",
    title = "{Merger and Mass Ejection of Neutron-Star Binaries}",
    eprint = "1908.02350",
    archivePrefix = "arXiv",
    primaryClass = "astro-ph.HE",
    doi = "10.1146/annurev-nucl-101918-023625",
    journal = "Ann. Rev. Nucl. Part. Sci.",
    volume = "69",
    pages = "41--64",
    year = "2019"
}

@article{Drout:2017ijr,
    author = "Drout, M. R. and others",
    title = "{Light Curves of the Neutron Star Merger GW170817/SSS17a: Implications for R-Process Nucleosynthesis}",
    eprint = "1710.05443",
    archivePrefix = "arXiv",
    primaryClass = "astro-ph.HE",
    doi = "10.1126/science.aaq0049",
    journal = "Science",
    volume = "358",
    pages = "1570--1574",
    year = "2017"
}

@article{LIGOScientific:2017ync,
    author = "Abbott, B. P. and others",
    collaboration = "LIGO Scientific, Virgo, Fermi GBM, INTEGRAL, IceCube, AstroSat Cadmium Zinc Telluride Imager Team, IPN, Insight-Hxmt, ANTARES, Swift, AGILE Team, 1M2H Team, Dark Energy Camera GW-EM, DES, DLT40, GRAWITA, Fermi-LAT, ATCA, ASKAP, Las Cumbres Observatory Group, OzGrav, DWF (Deeper Wider Faster Program), AST3, CAASTRO, VINROUGE, MASTER, J-GEM, GROWTH, JAGWAR, CaltechNRAO, TTU-NRAO, NuSTAR, Pan-STARRS, MAXI Team, TZAC Consortium, KU, Nordic Optical Telescope, ePESSTO, GROND, Texas Tech University, SALT Group, TOROS, BOOTES, MWA, CALET, IKI-GW Follow-up, H.E.S.S., LOFAR, LWA, HAWC, Pierre Auger, ALMA, Euro VLBI Team, Pi of Sky, Chandra Team at McGill University, DFN, ATLAS Telescopes, High Time Resolution Universe Survey, RIMAS, RATIR, SKA South Africa/MeerKAT",
    title = "{Multi-messenger Observations of a Binary Neutron Star Merger}",
    eprint = "1710.05833",
    archivePrefix = "arXiv",
    primaryClass = "astro-ph.HE",
    reportNumber = "LIGO-P1700294, VIR-0802A-17, FERMILAB-PUB-17-478-A-AE-CD",
    doi = "10.3847/2041-8213/aa91c9",
    journal = "Astrophys. J. Lett.",
    volume = "848",
    number = "2",
    pages = "L12",
    year = "2017"
}

@article{Metzger:2010sy,
    author = "Metzger, B. D. and Martinez-Pinedo, G. and Darbha, S. and Quataert, E. and Arcones, A. and Kasen, D. and Thomas, R. and Nugent, P. and Panov, I. V. and Zinner, N. T.",
    title = "{Electromagnetic Counterparts of Compact Object Mergers Powered by the Radioactive Decay of R-process Nuclei}",
    eprint = "1001.5029",
    archivePrefix = "arXiv",
    primaryClass = "astro-ph.HE",
    doi = "10.1111/j.1365-2966.2010.16864.x",
    journal = "Mon. Not. Roy. Astron. Soc.",
    volume = "406",
    pages = "2650",
    year = "2010"
}

@article{Lattimer:1977igd,
    author = "Lattimer, J. M. and Mackie, F. and Ravenhall, D. G. and Schramm, D. N.",
    title = "{The decompression of cold neutron star matter}",
    doi = "10.1086/155148",
    journal = "Astrophys. J.",
    volume = "213",
    pages = "225",
    year = "1977"
}

@article{Karpov2022,
  author  = {Karpov, P. I. and Huang, C. and Sitdikov, I. and Fryer, C. L. and Woosley, S. and Pilania, G.},
  title   = {{Physics-informed Machine Learning for Modeling Turbulence in Supernovae}},
  journal = {Astrophys. J.}, volume = {940}, number = {1}, pages = {26}, year = {2022},
  eprint  = {2205.08663}, archivePrefix = {arXiv}, doi = {10.3847/1538-4357/ac9c04}
}

@article{LIGOScientific:2017vwq,
    author = "Abbott, B. P. and others",
    collaboration = "LIGO Scientific, Virgo",
    title = "{GW170817: Observation of Gravitational Waves from a Binary Neutron Star Inspiral}",
    eprint = "1710.05832",
    archivePrefix = "arXiv",
    primaryClass = "gr-qc",
    reportNumber = "LIGO-P170817",
    doi = "10.1103/PhysRevLett.119.161101",
    journal = "Phys. Rev. Lett.",
    volume = "119",
    number = "16",
    pages = "161101",
    year = "2017"
}

@article{LIGOScientific:2017zic,
    author = "Abbott, B. P. and others",
    collaboration = "LIGO Scientific, Virgo, Fermi-GBM, INTEGRAL",
    title = "{Gravitational Waves and Gamma-rays from a Binary Neutron Star Merger: GW170817 and GRB 170817A}",
    eprint = "1710.05834",
    archivePrefix = "arXiv",
    primaryClass = "astro-ph.HE",
    reportNumber = "LIGO-P1700308",
    doi = "10.3847/2041-8213/aa920c",
    journal = "Astrophys. J. Lett.",
    volume = "848",
    number = "2",
    pages = "L13",
    year = "2017"
}

@article{Perego:2017wtu,
    author = "Perego, Albino and Radice, David and Bernuzzi, Sebastiano",
    title = "{AT 2017gfo: An Anisotropic and Three-component Kilonova Counterpart of GW170817}",
    eprint = "1711.03982",
    archivePrefix = "arXiv",
    primaryClass = "astro-ph.HE",
    doi = "10.3847/2041-8213/aa9ab9",
    journal = "Astrophys. J. Lett.",
    volume = "850",
    number = "2",
    pages = "L37",
    year = "2017"
}

@article{Kasen:2017sxr,
    author = "Kasen, Daniel and Metzger, Brian and Barnes, Jennifer and Quataert, Eliot and Ramirez-Ruiz, Enrico",
    title = "{Origin of the heavy elements in binary neutron-star mergers from a gravitational wave event}",
    eprint = "1710.05463",
    archivePrefix = "arXiv",
    primaryClass = "astro-ph.HE",
    doi = "10.1038/nature24453",
    journal = "Nature",
    volume = "551",
    pages = "80",
    year = "2017"
}

@article{Wanajo:2014wha,
    author = "Wanajo, Shinya and Sekiguchi, Yuichiro and Nishimura, Nobuya and Kiuchi, Kenta and Kyutoku, Koutarou and Shibata, Masaru",
    title = "{Production of all the $r$-process nuclides in the dynamical ejecta of neutron star mergers}",
    eprint = "1402.7317",
    archivePrefix = "arXiv",
    primaryClass = "astro-ph.SR",
    doi = "10.1088/2041-8205/789/2/L39",
    journal = "Astrophys. J. Lett.",
    volume = "789",
    pages = "L39",
    year = "2014"
}

@article{Radice:2018pdn,
    author = "Radice, David and Perego, Albino and Hotokezaka, Kenta and Fromm, Steven A. and Bernuzzi, Sebastiano and Roberts, Luke F.",
    title = "{Binary Neutron Star Mergers: Mass Ejection, Electromagnetic Counterparts and Nucleosynthesis}",
    eprint = "1809.11161",
    archivePrefix = "arXiv",
    primaryClass = "astro-ph.HE",
    doi = "10.3847/1538-4357/aaf054",
    journal = "Astrophys. J.",
    volume = "869",
    number = "2",
    pages = "130",
    year = "2018"
}

@article{Radice:2020ddv,
    author = "Radice, David and Bernuzzi, Sebastiano and Perego, Albino",
    title = "{The Dynamics of Binary Neutron Star Mergers and GW170817}",
    eprint = "2002.03863",
    archivePrefix = "arXiv",
    primaryClass = "astro-ph.HE",
    doi = "10.1146/annurev-nucl-013120-114541",
    journal = "Ann. Rev. Nucl. Part. Sci.",
    volume = "70",
    pages = "95--119",
    year = "2020"
}

@article{Qian:1996xt,
    author = "Qian, Y. Z. and Woosley, S. E.",
    title = "{Nucleosynthesis in neutrino driven winds: 1. The Physical conditions}",
    eprint = "astro-ph/9611094",
    archivePrefix = "arXiv",
    doi = "10.1086/177973",
    journal = "Astrophys. J.",
    volume = "471",
    pages = "331--351",
    year = "1996"
}

@article{Hoffman:1996aj,
    author = "Hoffman, R. D. and Woosley, S. E. and Qian, Y. Z.",
    title = "{Nucleosynthesis in neutrino driven winds: 2. Implications for heavy element synthesis}",
    eprint = "astro-ph/9611097",
    archivePrefix = "arXiv",
    doi = "10.1086/304181",
    journal = "Astrophys. J.",
    volume = "482",
    pages = "951",
    year = "1997"
}

@article{Martin:2017dhc,
    author = "Martin, D. and Perego, A. and Kastaun, W. and Arcones, A.",
    title = "{The role of weak interactions in dynamic ejecta from binary neutron star mergers}",
    eprint = "1710.04900",
    archivePrefix = "arXiv",
    primaryClass = "astro-ph.HE",
    doi = "10.1088/1361-6382/aa9f5a",
    journal = "Class. Quant. Grav.",
    volume = "35",
    number = "3",
    pages = "034001",
    year = "2018"
}

@article{Shibata:2011kx,
    author = "Shibata, Masaru and Kiuchi, Kenta and Sekiguchi, Yu-ichiro and Suwa, Yudai",
    title = "{Truncated Moment Formalism for Radiation Hydrodynamics in Numerical Relativity}",
    eprint = "1104.3937",
    archivePrefix = "arXiv",
    primaryClass = "astro-ph.HE",
    doi = "10.1143/PTP.125.1255",
    journal = "Prog. Theor. Phys.",
    volume = "125",
    pages = "1255--1287",
    year = "2011"
}

@article{Radice:2021jtw,
    author = "Radice, David and Bernuzzi, Sebastiano and Perego, Albino and Haas, Roland",
    title = "{A new moment-based general-relativistic neutrino-radiation transport code: Methods and first applications to neutron star mergers}",
    eprint = "2111.14858",
    archivePrefix = "arXiv",
    primaryClass = "astro-ph.HE",
    doi = "10.1093/mnras/stac589",
    journal = "Mon. Not. Roy. Astron. Soc.",
    volume = "512",
    number = "1",
    pages = "1499--1521",
    year = "2022"
}

@article{Foucart:2022bth,
    author = "Foucart, Francois",
    title = "{Neutrino transport in general relativistic neutron star merger simulations}",
    eprint = "2209.02538",
    archivePrefix = "arXiv",
    primaryClass = "astro-ph.HE",
    doi = "10.1007/s41115-023-00016-y",
    journal = "Liv. Rev. Comput. Astrophys.",
    volume = "9",
    number = "1",
    pages = "1",
    year = "2023"
}

@article{Sekiguchi:2015dma,
    author = "Sekiguchi, Yuichiro and Kiuchi, Kenta and Kyutoku, Koutarou and Shibata, Masaru",
    title = "{Dynamical mass ejection from binary neutron star mergers: Radiation-hydrodynamics study in general relativity}",
    eprint = "1502.06660",
    archivePrefix = "arXiv",
    primaryClass = "astro-ph.HE",
    doi = "10.1103/PhysRevD.91.064059",
    journal = "Phys. Rev. D",
    volume = "91",
    number = "6",
    pages = "064059",
    year = "2015"
}

@article{Foucart:2016rxm,
    author = "Foucart, Francois and O'Connor, Evan and Roberts, Luke and Kidder, Lawrence E. and Pfeiffer, Harald P. and Scheel, Mark A.",
    title = "{Impact of an improved neutrino energy estimate on outflows in neutron star merger simulations}",
    eprint = "1607.07450",
    archivePrefix = "arXiv",
    primaryClass = "astro-ph.HE",
    doi = "10.1103/PhysRevD.94.123016",
    journal = "Phys. Rev. D",
    volume = "94",
    number = "12",
    pages = "123016",
    year = "2016"
}

@article{Chiesa:2024lnu,
    author = "Chiesa, Leonardo and Bhattacharyya, Maitraya and Mazzini, Filippo and Guercilena, Federico Maria and Perego, Albino and Radice, David",
    title = "{Open-source library for performance-portable neutrino reaction rates: Application to neutron star mergers}",
    eprint = "2412.04570",
    archivePrefix = "arXiv",
    primaryClass = "astro-ph.HE",
    doi = "10.1103/PhysRevD.111.063053",
    journal = "Phys. Rev. D",
    volume = "111",
    number = "6",
    pages = "063053",
    year = "2025"
}

@article{Cheong:2024cnb,
    author = "Cheong, Patrick Chi-Kit and Foucart, Francois and Ng, Harry Ho-Yin and Offermans, Arthur and Duez, Matthew D. and Muhammed, Nishad and Chawhan, Pavan",
    title = "{Influence of neutrino-electron scattering and neutrino-pair annihilation on hypermassive neutron star}",
    eprint = "2410.20681",
    archivePrefix = "arXiv",
    primaryClass = "astro-ph.HE",
    doi = "10.1103/PhysRevD.111.043036",
    journal = "Phys. Rev. D",
    volume = "111",
    number = "4",
    pages = "043036",
    year = "2025"
}

@article{Typel:2009sy,
    author = "Typel, S. and Ropke, G. and Klahn, T. and Blaschke, D. and Wolter, H. H.",
    title = "{Composition and thermodynamics of nuclear matter with light clusters}",
    eprint = "0908.2344",
    archivePrefix = "arXiv",
    primaryClass = "nucl-th",
    doi = "10.1103/PhysRevC.81.015803",
    journal = "Phys. Rev. C",
    volume = "81",
    pages = "015803",
    year = "2010"
}

@article{Hempel:2009mc,
    author = "Hempel, Matthias and Schaffner-Bielich, Jurgen",
    title = "{Statistical Model for a Complete Supernova Equation of State}",
    eprint = "0911.4073",
    archivePrefix = "arXiv",
    primaryClass = "nucl-th",
    doi = "10.1016/j.nuclphysa.2010.02.010",
    journal = "Nucl. Phys. A",
    volume = "837",
    pages = "210--254",
    year = "2010"
}

@article{Gourgoulhon:2000nn,
    author = "Gourgoulhon, Eric and Grandclement, Philippe and Taniguchi, Keisuke and Marck, Jean-Alain and Bonazzola, Silvano",
    title = "{Quasiequilibrium sequences of synchronized and irrotational binary neutron stars in general relativity: 1. Method and tests}",
    eprint = "gr-qc/0007028",
    archivePrefix = "arXiv",
    doi = "10.1103/PhysRevD.63.064029",
    journal = "Phys. Rev. D",
    volume = "63",
    pages = "064029",
    year = "2001"
}

@article{Berger:1984zza,
    author = "Berger, Marsha J. and Oliger, Joseph",
    title = "{Adaptive Mesh Refinement for Hyperbolic Partial Differential Equations}",
    doi = "10.1016/0021-9991(84)90073-1",
    journal = "J. Comput. Phys.",
    volume = "53",
    pages = "484",
    year = "1984"
}

@Article{Berger:1989a,
  doi		= {10.1016/0021-9991(89)90035-1},
  url		= {http://dx.doi.org/10.1016/0021-9991(89)90035-1},
  year		= {1989},
  month		= {may},
  publisher	= {Elsevier {BV}},
  volume	= {82},
  number	= {1},
  pages		= {64--84},
  author	= {M.J. Berger and P. Colella},
  title		= {Local adaptive mesh refinement for shock hydrodynamics},
  journal	= {J. Comput. Phys.}
}

@article{Qiu:2025kgy,
    author = "Qiu, Yi and Radice, David and Richers, Sherwood and Bhattacharyya, Maitraya",
    title = "{Neutrino Flavor Transformation in Neutron Star Mergers}",
    eprint = "2503.11758",
    archivePrefix = "arXiv",
    primaryClass = "astro-ph.HE",
    reportNumber = "INT-PUB-25-004",
    doi = "10.1103/h2q7-kn3v",
    journal = "Phys. Rev. Lett.",
    volume = "135",
    number = "9",
    pages = "091401",
    year = "2025"
}

@article{Qiu:2025ybw,
    author = "Qiu, Yi and Radice, David and Richers, Sherwood and Guercilena, Federico Maria and Perego, Albino and Bhattacharyya, Maitraya",
    title = "{Impact of neutrino flavor conversions on neutron star merger dynamics, ejecta, nucleosynthesis, and multimessenger signals}",
    eprint = "2510.15028",
    archivePrefix = "arXiv",
    primaryClass = "astro-ph.HE",
    doi = "10.1103/qckq-78gt",
    journal = "Phys. Rev. D",
    volume = "112",
    number = "12",
    pages = "123039",
    year = "2025"
}

@article{Baiotti:2016qnr,
    author = "Baiotti, Luca and Rezzolla, Luciano",
    title = "{Binary neutron star mergers: a review of Einstein\textquoteright{}s richest laboratory}",
    eprint = "1607.03540",
    archivePrefix = "arXiv",
    primaryClass = "gr-qc",
    doi = "10.1088/1361-6633/aa67bb",
    journal = "Rept. Prog. Phys.",
    volume = "80",
    number = "9",
    pages = "096901",
    year = "2017"
}

@article{Thorne:1981nvt,
    author = "Thorne, Kip S.",
    title = "{Relativistic radiative transfer: moment formalisms}",
    doi = "10.1093/mnras/194.2.439",
    journal = "Mon. Not. Roy. Astron. Soc.",
    volume = "194",
    number = "2",
    pages = "439--473",
    year = "1981"
}

@article{Fields:2025uyk,
    author = "Fields, Jacob and Radice, David",
    title = "{AthenaK Simulations of Magnetized Binary Neutron Star Mergers}",
    eprint = "2511.06058",
    archivePrefix = "arXiv",
    primaryClass = "astro-ph.HE",
    doi = "10.1088/1742-6596/3252/1/012003",
    journal = "J. Phys. Conf. Ser.",
    volume = "3252",
    number = "1",
    pages = "012003",
    year = "2026"
}

@article{Cardall:2013kwa,
    author = "Cardall, Christian Y. and Endeve, Eirik and Mezzacappa, Anthony",
    title = "{Conservative 3+1 General Relativistic Boltzmann Equation}",
    eprint = "1305.0037",
    archivePrefix = "arXiv",
    primaryClass = "astro-ph.HE",
    doi = "10.1103/PhysRevD.88.023011",
    journal = "Phys. Rev. D",
    volume = "88",
    pages = "023011",
    year = "2013"
}

@article{Cusinato:2021zin,
    author = "Cusinato, Marco and Guercilena, Federico Maria and Perego, Albino and Logoteta, Domenico and Radice, David and Bernuzzi, Sebastiano and Ansoldi, Stefano",
    title = "{Neutrino emission from binary neutron star mergers: characterizing light curves and mean energies}",
    eprint = "2111.13005",
    archivePrefix = "arXiv",
    primaryClass = "astro-ph.HE",
    doi = "10.1140/epja/s10050-022-00743-5",
    month = "11",
    year = "2021"
}

@article{Thummler:2026baz,
    author = {Th{\"u}mmler, L. and Kuroda, T.},
    title = "{When a neural surrogate cannot accelerate a solver: runtime share, closed-loop drift, and the economics of uncertainty gating in a stiff coupled simulation}",
    eprint = "2608.23075",
    archivePrefix = "arXiv",
    primaryClass = "astro-ph.IM",
    month = "8",
    year = "2026"
}

@article{Cheong:2024buu,
    author = "Cheong, Patrick Chi-Kit and Foucart, Francois and Duez, Matthew D. and Offermans, Arthur and Muhammed, Nishad and Chawhan, Pavan",
    title = "{Energy-dependent and Energy-integrated Two-moment General-relativistic Neutrino Transport Simulations of a Hypermassive Neutron Star}",
    eprint = "2407.16017",
    archivePrefix = "arXiv",
    primaryClass = "astro-ph.HE",
    doi = "10.3847/1538-4357/ad7825",
    journal = "Astrophys. J.",
    volume = "975",
    number = "1",
    pages = "116",
    year = "2024"
}

@article{Foucart:2024npn,
    author = "Foucart, Francois and Cheong, Patrick Chi-Kit and Duez, Matthew D. and Kidder, Lawrence E. and Pfeiffer, Harald P. and Scheel, Mark A.",
    title = "{Robustness of neutron star merger simulations to changes in neutrino transport and neutrino-matter interactions}",
    eprint = "2407.15989",
    archivePrefix = "arXiv",
    primaryClass = "astro-ph.HE",
    doi = "10.1103/PhysRevD.110.083028",
    journal = "Phys. Rev. D",
    volume = "110",
    number = "8",
    pages = "083028",
    year = "2024"
}

@article{Sekiguchi:2016bjd,
    author = "Sekiguchi, Yuichiro and Kiuchi, Kenta and Kyutoku, Koutarou and Shibata, Masaru and Taniguchi, Keisuke",
    title = "{Dynamical mass ejection from the merger of asymmetric binary neutron stars: Radiation-hydrodynamics study in general relativity}",
    eprint = "1603.01918",
    archivePrefix = "arXiv",
    primaryClass = "astro-ph.HE",
    doi = "10.1103/PhysRevD.93.124046",
    journal = "Phys. Rev. D",
    volume = "93",
    number = "12",
    pages = "124046",
    year = "2016"
}

@article{Foucart:2017mbt,
    author = "Foucart, Francois",
    title = "{Monte Carlo closure for moment-based transport schemes in general relativistic radiation hydrodynamic simulations}",
    eprint = "1708.08452",
    archivePrefix = "arXiv",
    primaryClass = "astro-ph.HE",
    doi = "10.1093/mnras/sty108",
    journal = "Mon. Not. Roy. Astron. Soc.",
    volume = "475",
    number = "3",
    pages = "4186--4207",
    year = "2018"
}

@article{Wu:2016pnw,
    author = "Wu, Meng-Ru and Fern\'andez, Rodrigo and Mart\'\i{}nez-Pinedo, Gabriel and Metzger, Brian D.",
    title = "{Production of the entire range of r-process nuclides by black hole accretion disc outflows from neutron star mergers}",
    eprint = "1607.05290",
    archivePrefix = "arXiv",
    primaryClass = "astro-ph.HE",
    doi = "10.1093/mnras/stw2156",
    journal = "Mon. Not. Roy. Astron. Soc.",
    volume = "463",
    number = "3",
    pages = "2323--2334",
    year = "2016"
}

@article{Pajkos:2024iry,
    author = "Pajkos, Michael A. and Most, Elias R.",
    title = "{Influence of muons, pions, and trapped neutrinos on neutron star mergers}",
    eprint = "2409.09147",
    archivePrefix = "arXiv",
    primaryClass = "astro-ph.HE",
    doi = "10.1103/PhysRevD.111.043013",
    journal = "Phys. Rev. D",
    volume = "111",
    number = "4",
    pages = "043013",
    year = "2025"
}

@article{Ng:2024zve,
    author = "Ng, Harry Ho-Yin and Musolino, Carlo and Tootle, Samuel D. and Rezzolla, Luciano",
    title = "{Accurate Muonic Interactions in Neutron Star Mergers and Impact on Heavy-element Nucleosynthesis}",
    eprint = "2411.19178",
    archivePrefix = "arXiv",
    primaryClass = "astro-ph.HE",
    doi = "10.3847/2041-8213/add324",
    journal = "Astrophys. J. Lett.",
    volume = "985",
    number = "2",
    pages = "L36",
    year = "2025"
}

@article{Ng:2023syk,
    author = "Ng, Harry Ho-Yin and Cheong, Patrick Chi-Kit and Lam, Alan Tsz-Lok and Li, Tjonnie Guang Feng",
    title = "{General-relativistic Radiation Transport Scheme in Gmunu. II. Implementation of Novel Microphysical Library for Neutrino Radiation{\textemdash}Weakhub}",
    eprint = "2309.03526",
    archivePrefix = "arXiv",
    primaryClass = "astro-ph.HE",
    doi = "10.3847/1538-4365/ad2fbd",
    journal = "Astrophys. J. Suppl.",
    volume = "272",
    number = "1",
    pages = "9",
    year = "2024"
}

@article{Foucart:2026kpd,
    author = "Foucart, Francois and Rath, Samantha and Davidson, Rowan and Cheong, Patrick Chi-Kit and Duez, Matthew D. and Kidder, Lawrence and Pfeiffer, Harald and Scheel, Mark",
    title = "{Impact of neutrino-electron scattering and an improved treatment of pair processes on binary neutron star mergers}",
    eprint = "2606.27425",
    archivePrefix = "arXiv",
    primaryClass = "astro-ph.HE",
    month = "6",
    year = "2026"
}

@article{Fields:2024pob,
    author = "Fields, Jacob and Zhu, Hengrui and Radice, David and Stone, James M. and Cook, William and Bernuzzi, Sebastiano and Daszuta, Boris",
    title = "{Performance-portable Binary Neutron Star Mergers with AthenaK}",
    eprint = "2409.10384",
    archivePrefix = "arXiv",
    primaryClass = "astro-ph.HE",
    doi = "10.3847/1538-4365/ad9687",
    journal = "Astrophys. J. Suppl.",
    volume = "276",
    number = "2",
    pages = "35",
    year = "2025"
}

@inproceedings{Kingma:2014adam,
    author = "Kingma, Diederik P. and Ba, Jimmy",
    title = "{Adam: A Method for Stochastic Optimization}",
    booktitle = "{International Conference on Learning Representations}",
    eprint = "1412.6980",
    archivePrefix = "arXiv",
    primaryClass = "cs.LG",
    month = "12",
    year = "2014"
}

@article{Izquierdo:2023fub,
    author = "Izquierdo, Manuel R. and Abalos, J. Fernando and Palenzuela, Carlos",
    title = "{Guided moments formalism: A new efficient full-neutrino treatment for astrophysical simulations}",
    eprint = "2312.09275",
    archivePrefix = "arXiv",
    primaryClass = "astro-ph.HE",
    doi = "10.1103/PhysRevD.109.043044",
    journal = "Phys. Rev. D",
    volume = "109",
    number = "4",
    pages = "043044",
    year = "2024"
}

@article{KokkosCore2022,
	title        = {Kokkos 3: Programming Model Extensions for the Exascale Era},
	author       = {
		Trott, Christian R. and Lebrun-Grandié, Damien and Arndt, Daniel and Ciesko,
		Jan and Dang, Vinh and Ellingwood, Nathan and Gayatri, Rahulkumar and Harvey,
		Evan and Hollman, Daisy S. and Ibanez, Dan and Liber, Nevin and Madsen,
		Jonathan and Miles, Jeff and Poliakoff, David and Powell, Amy and
		Rajamanickam, Sivasankaran and Simberg, Mikael and Sunderland, Dan and
		Turcksin, Bruno and Wilke, Jeremiah
	},
	year         = 2022,
	journal      = {IEEE Transactions on Parallel and Distributed Systems},
	volume       = 33,
	number       = 4,
	pages        = {805--817},
	doi          = {10.1109/TPDS.2021.3097283}
}

@inproceedings{KokkosTools2018,
	title        = {Profiling and Debugging Support for the Kokkos Programming Model},
	author       = {
		Hammond, Simon D. and Trott, Christian R. and Ibanez, Daniel and Sunderland,
		Daniel
	},
	year         = 2018,
	booktitle    = {High Performance Computing},
	publisher    = {Springer International Publishing},
	address      = {Cham},
	pages        = {743--754},
	doi          = {10.1007/978-3-030-02465-9_53},
	isbn         = {978-3-030-02465-9}
}

@article{Stone_2026,
doi = {10.3847/1538-4365/ae3717},
url = {https://doi.org/10.3847/1538-4365/ae3717},
year = {2026},
month = {feb},
publisher = {The American Astronomical Society},
volume = {283},
number = {1},
pages = {27},
author = {Stone, James M. and Mullen, Patrick D. and Fielding, Drummond and Grete, Philipp and Guo, Minghao and Kempski, Philipp and Most, Elias R. and White, Christopher J. and Wong, George N.},
title = {AthenaK: A Performance-portable Version of the Athena++ Adaptive Mesh Refinement Framework},
journal = {The Astrophysical Journal Supplement Series}
}

@article{Zhu:2024utz,
    author = "Zhu, Hengrui and Fields, Jacob and Zappa, Francesco and Radice, David and Stone, James M. and Rashti, Alireza and Cook, William and Bernuzzi, Sebastiano and Daszuta, Boris",
    title = "{Performance-portable Numerical Relativity with AthenaK}",
    eprint = "2409.10383",
    archivePrefix = "arXiv",
    primaryClass = "gr-qc",
    doi = "10.3847/1538-4365/adcf96",
    journal = "Astrophys. J. Suppl.",
    volume = "278",
    number = "2",
    pages = "50",
    year = "2025"
}

@article{Rath:2026vfr,
    author = "Rath, Samantha and Foucart, Francois and Kidder, Lawrence E. and Pfeiffer, Harald P. and Scheel, Mark A.",
    title = "{Assessing the Relative Importance of Neutrino Matter Interaction Channels in Post-Merger Remnant of Binary Neutron Stars}",
    eprint = "2605.29187",
    archivePrefix = "arXiv",
    primaryClass = "astro-ph.HE",
    month = "5",
    year = "2026"
}

@article{OConnor:2014sgn,
    author = "O'Connor, Evan",
    title = "{An Open-Source Neutrino Radiation Hydrodynamics Code for Core-Collapse Supernovae}",
    eprint = "1411.7058",
    archivePrefix = "arXiv",
    primaryClass = "astro-ph.HE",
    doi = "10.1088/0067-0049/219/2/24",
    journal = "Astrophys. J. Suppl.",
    volume = "219",
    number = "2",
    pages = "24",
    year = "2015"
}

@article{Gieg:2026beb,
    author = "Gieg, Henrique and Jaeger, Ramon and Ujevic, Maximiliano and Dietrich, Tim",
    title = "{Consistent treatment of muons in binary neutron star mergers}",
    eprint = "2604.14225",
    archivePrefix = "arXiv",
    primaryClass = "astro-ph.HE",
    doi = "10.1103/qmb8-q3cx",
    journal = "Phys. Rev. D",
    volume = "114",
    number = "4",
    pages = "043036",
    year = "2026"
}

@article{Gieg:2024jxs,
    author = "Gieg, Henrique and Schianchi, Federico and Ujevic, Maximiliano and Dietrich, Tim",
    title = "{Role of muons in binary neutron star mergers: First simulations}",
    eprint = "2409.04420",
    archivePrefix = "arXiv",
    primaryClass = "gr-qc",
    doi = "10.1103/52ph-53yw",
    journal = "Phys. Rev. D",
    volume = "112",
    number = "2",
    pages = "023036",
    year = "2025"
}

@article{Richers:2024zit,
    author = "Richers, Sherwood and Froustey, Julien and Ghosh, Somdutta and Foucart, Francois and Gomez, Javier",
    title = "{Asymptotic-state prediction for fast flavor transformation in neutron star mergers}",
    eprint = "2409.04405",
    archivePrefix = "arXiv",
    primaryClass = "astro-ph.HE",
    reportNumber = "N3AS-24-030",
    doi = "10.1103/PhysRevD.110.103019",
    journal = "Phys. Rev. D",
    volume = "110",
    number = "10",
    pages = "103019",
    year = "2024"
}

@article{Abbar:2023kta,
    author = "Abbar, Sajad",
    title = "{Applications of machine learning to detecting fast neutrino flavor instabilities in core-collapse supernova and neutron star merger models}",
    eprint = "2303.05560",
    archivePrefix = "arXiv",
    primaryClass = "astro-ph.HE",
    reportNumber = "MPP-2023-39",
    doi = "10.1103/PhysRevD.107.103006",
    journal = "Phys. Rev. D",
    volume = "107",
    number = "10",
    pages = "103006",
    year = "2023"
}

@article{Chakraborty:2026vhk,
    author = "Chakraborty, Madhurima and Abbar, Sajad and Wu, Meng-Ru and Foucart, Francois and Xiong, Zewei",
    title = "{Machine Learning Detection of Non-Axisymmetric Fast Flavor Instabilities in Compact Objects}",
    eprint = "2607.12558",
    archivePrefix = "arXiv",
    primaryClass = "astro-ph.HE",
    month = "7",
    year = "2026"
}

@article{Cook:2026tum,
    author = "Cook, William and Bernuzzi, Sebastiano",
    title = "{Subgrid Modelling for Relativistic Magnetohydrodynamics with Machine Learning}",
    eprint = "2606.21659",
    archivePrefix = "arXiv",
    primaryClass = "astro-ph.HE",
    month = "6",
    year = "2026"
}

@inproceedings{Bhojwani:2025rtn,
    author = "Bhojwani, Nihaal and Wang, Chuwei and Wang, Hai-Yang and Sun, Chang and Most, Elias R. and Anandkumar, Anima",
    title = "{From Black Hole to Galaxy: Neural Operator: Framework for Accretion and Feedback Dynamics}",
    booktitle = "{38th conference on Neural Information Processing Systems}",
    eprint = "2512.01576",
    archivePrefix = "arXiv",
    primaryClass = "astro-ph.HE",
    month = "12",
    year = "2025"
}

@article{Hendrycks:2016gelu,
    author = "Hendrycks, Dan and Gimpel, Kevin",
    title = "{Gaussian Error Linear Units (GELUs)}",
    eprint = "1606.08415",
    archivePrefix = "arXiv",
    primaryClass = "cs.LG",
    month = "6",
    year = "2016"
}

@article{He:2016resnet,
    author = "He, Kaiming and Zhang, Xiangyu and Ren, Shaoqing and Sun, Jian",
    title = "{Deep Residual Learning for Image Recognition}",
    eprint = "1512.03385",
    archivePrefix = "arXiv",
    primaryClass = "cs.CV",
    doi = "10.1109/CVPR.2016.90",
    month = "12",
    year = "2015"
}

@article{Ba:2016layernorm,
    author = "Ba, Jimmy Lei and Kiros, Jamie Ryan and Hinton, Geoffrey E.",
    title = "{Layer Normalization}",
    eprint = "1607.06450",
    archivePrefix = "arXiv",
    primaryClass = "stat.ML",
    month = "7",
    year = "2016"
}

@article{Paszke:2019xhz,
    author = "Paszke, Adam and others",
    title = "{PyTorch: An Imperative Style, High-Performance Deep Learning Library}",
    eprint = "1912.01703",
    archivePrefix = "arXiv",
    primaryClass = "cs.LG",
    month = "12",
    year = "2019"
}

@article{CompOSE2022,
    author = "Typel, S. and others",
    collaboration = "CompOSE Core Team",
    title = "{CompOSE Reference Manual}",
    eprint = "2203.03209",
    archivePrefix = "arXiv",
    primaryClass = "astro-ph.HE",
    doi = "10.1140/epja/s10050-022-00847-y",
    journal = "Eur. Phys. J. A",
    volume = "58",
    number = "11",
    pages = "221",
    year = "2022"
}

@Inbook{LeCun:1998,
author="LeCun, Yann
and Bottou, Leon
and Orr, Genevieve B.
and M{\"u}ller, Klaus -Robert",
editor="Orr, Genevieve B.
and M{\"u}ller, Klaus-Robert",
title="Efficient BackProp",
bookTitle="Neural Networks: Tricks of the Trade",
year="1998",
publisher="Springer Berlin Heidelberg",
address="Berlin, Heidelberg",
pages="9--50",
isbn="978-3-540-49430-0",
doi="10.1007/3-540-49430-8_2",
url="https://doi.org/10.1007/3-540-49430-8_2"
}
%%%%%%%%%%%%%%%%%%%%%%%%%%%%%%%%%%%%%%%%%%%%%%%%%%%%%%%%%%%%%%%%%%%%%

\end{document}